\documentclass[%
 reprint,
 amsmath,amssymb,
 aps
]{revtex4-2}

\usepackage{graphicx}% Include figure files
\usepackage{dcolumn}% Align table columns on 
\usepackage{xcolor} 
\usepackage{bm}% bold math
\usepackage{comment}
\usepackage{float}

\usepackage{hyperref}% add hypertext capabilities
\begin{document}

%\preprint{APS/123-QED}

%\title{Precessional Dynamics of Electromagnetic Spin in Structured Fields: Toward Photonic Spin Waves}
\title{Optical spin precession}

% Force line breaks with \\

\author{Abanoub Mikhail$^{1,2}$}
\email{These authors have equally contributed to this work}

\author{Maxim Mazanov$^{1,2}$}
\email{These authors have equally contributed to this work}

\author{Ilya Deiry$^{1,2}$}

\author{Mingzhao Song$^{1}$}
\email{kevinsmz@foxmail.com}

\author{Ivan Iorsh$^{3}$}

\author{Andrey Bogdanov$^{1,2}$}
\email{a.bogdanov@hrbeu.edu.cn}

\affiliation{$^{1}$ Qingdao Innovation and Development Center of Harbin Engineering University, 266404, Qingdao, China}
\affiliation{$^{2}$ Faculty of Physics and Engineering, ITMO University, Saint Petersburg, Russia}
\affiliation{$^{3}$ Department of Physics, Engineering Physics $\&$ Astronomy, Queen’s Universiy, Kingston, Canada}

%\affiliation{%
%$^1$Faculty of Physics and Engineering, ITMO University, Saint Petersburg, Russia\\
%$^2$Qingdao Innovation and Development Center, Harbin Engineering University, %266404, Qingdao, China
%}

\date{\today}% It is always \today, today,

\begin{abstract}
Period-averaged electromagnetic spin angular momentum is a well-established quantity for monochromatic fields, governing phenomena such as light–matter interactions with chiral particles and spin-orbit coupling effects. In contrast, the spin angular momentum of  non-monochromatic fields remains unexplored.  Here, we extend the concept of optical spin to the domain of non-monochromatic electromagnetic fields. Through this formulation, we uncover the precessional dynamics of electromagnetic spin in specific polychromatic configurations, including the superposition of circularly and linearly polarized plane waves propagating orthogonally at different frequencies, as well as fields generated by a precessing magnetic dipole. We discover that the dynamics of the electromagnetic spin in these cases obeys a Landau-Lifshitz-like equation establishing a profound parallel between dynamics of magnetization and photonic spin. 
\end{abstract}

\maketitle

%\tableofcontents

\section{\label{sec:level1} Introduction}

Electromagnetic fields carry energy, momentum, and angular momentum. The total angular momentum of light can be decomposed into orbital and spin contributions~\cite{Jackson1999,Boyer1984}. The total angular momentum operator generates rotations for any three-dimensional vector field. The orbital component rotates the field's spatial distribution, whereas the spin component rotates the field direction locally~\cite{Bliokh2025}.  

Another possible dynamics for electromagnetic fields is %\emph{
%\textcolor{red}{spin} 
\textit{precession}, %}, 
which combines a rotation of the two-dimensional field projections with a constant third component, mimicking %The common example being 
precession of spinning tops and gyroscopes~\cite{Goldstein200}.  The same governing principle of precessional motion appears in condensed matter, where the magnetization precesses about an effective magnetic field %and relaxes through damping 
as described phenomenologically by Landau-Lifshitz equation~\cite{Gilbert2004,Rozkotova}. Larmor precession of nuclear and electronic magnetic moments in an applied external magnetic field underpins magnetic-resonance phenomena and is described by the Bloch equations; this behavior is exploited in nuclear magnetic resonance and electron paramagnetic resonance~\cite{Bloch1946,Hall1964,Weil2007}. 
%\textcolor{red}{maybe this paragraph is a bit too defensive. need to reconsider} - \textcolor{green}{REVISED}
%A natural question is if electromagnetic fields could possess 

The electromagnetic \textit{spin angular momentum} (SAM)  is a well-established quantity for monochromatic waves,  usually described by its space-dependent time average~\cite{Bliokh2025}. While the spin of circular-polarized plane waves is longitudinal~\cite{Cameron2012,Van1993,Lenstra1982}, recent studies have revealed the transverse spin ubiquitous for localized waves such as surface polaritons~\cite{Bliokh2012Jun,Neugebauer2015Feb,Shi2023,Yermakov2016} and even in two-wave interference~\cite{Bekshaev2015Mar}. However, for free monochromatic fields the spin angular momentum is conserved, $\dot{\bm{S}} = 0$, and only interaction with matter can alter it~\cite{Van1993}. 

%In contrast to time-averaged electromagnetic quantities, instantaneous quantities depends on the exact values of electric and magnetic fields at any given time, encompassing all transient and non-periodic behaviors. This distinction is crucial, as time-average assumptions often lead different results from the instantaneous ones \cite{Mirmoosa2020}. Most of previous studies focused on the spin of harmonically time-varying optical fields \cite{Bliokh2017,Shi2023}. The dynamics of instantaneous spin of non-monochromatic waves, however, remains largely unexplored.    usually described by its space-dependent time average~\cite{Bliokh2025}. 

In classical mechanics and spintronics, spin precession is a widely studied phenomenon; yet, an optical analog of this effect has remained an open question ~\cite{Zvezdin2022,Awschalom2013}. Previous studies of optical spin have primarily focused on monochromatic fields and on systems without sources~\cite{Bliokh2017,Shi2023,Karakhanyan2024Jul}. Consequently, the SAM and its dynamics in polychromatic fields, particularly in the presence of sources, remain largely unexplored. It is commonly known that spatial inhomogeneity can generate non-trivial spatial configurations of optical spin, known as optical skyrmions~\cite{Shen2024,Tsesses2018,Shen2022}. Here, we show that temporal inhomogeneity (a polychromatic field) instead leads to non-trivial spin dynamics, such as optical spin precession.

%In this work, 
We extend the concept of optical (electromagnetic) spin beyond the monochromatic domain and demonstrate that it can undergo genuine precessional motion, analogous to the precession of elementary particles and magnetic moments in external magnetic fields. This optical spin precession is a distinctly different phenomenon from polarization rotation or other space-dependent polarization effects. 
For instance, gyrotropy describes asymmetric scattering of fields with opposite circular polarizations, leading to spatial rotation of the polarization plane (optical activity), while circular birefringence can originate from either internal chirality or external magnetic bias~\cite{Carey2024,Iguchi2021}. Likewise, precession of Stokes vectors along coiled ray trajectories has been studied in quasi-isotropic geometrical optics, but only for monochromatic fields~\cite{Bliokh2008}. Polarization rotation arising from Berry-phase effects is also well established ~\cite{Gangaraj2017}. In contrast, the precession we reveal is intrinsic to non-monochromatic electromagnetic fields and represents a new form of spin dynamics.

The rest of this paper is organized as follows. In Section~\ref{sec_theoretical_framework}, we employ the Belinfante spin-density formalism to derive a gauge-invariant expression for the spin density of non-monochromatic fields with magnetic sources, together with its electromagnetic dual for electric sources. From this formulation, we obtain a modified continuity equation that includes a spin-source term localized at the origin — a spin-torque density proportional to $-\mathbf{M}\times\mathbf{B}$, where $\mathbf{M}$ is the magnetization density (Section~\ref{subsec_spin_density_sources_continuity_equation}). Motivated by the magnetic-field structure encountered in ferromagnetic resonance, Section~\ref{subsec_circualrly_polarized_wave_constant_field} demonstrates that a circularly polarized plane wave interacting with a longitudinal magnetic field exhibits precession of the optical spin. To bring the precession frequency into an experimentally accessible range, Section~\ref{subsec_two_wave_interference} considers a generalization: a pair of circularly and linearly polarized plane waves propagating orthogonally at different frequencies $\omega_{1,2}$, producing precession at the difference frequency $\delta \omega = \omega_2 - \omega_1$ and nutation at the sum frequency $\omega_1 + \omega_2$. In the quasi-monochromatic limit $\delta\omega\ll\omega_1+\omega_2$, low-frequency precession dominates and nutation is negligible. Section~\ref{sec_precessing_magnetic_dipole} addresses the opposite perspective of ferromagnetic resonance, where we consider the field generated by a precessing magnetic dipole. In the near field, the local electromagnetic spin density follows synchronously with the dipole precession, while the total spin precesses at the same frequency. Finally, Section~\ref{sec_experiments} outlines two experimental realizations: a quasi-monochromatic two-beam interference scheme and a microwave implementation using three-dimensional split-ring meta-atoms with independently controlled currents.

\section{Instantaneous spin density}\label{sec_theoretical_framework}
%\noindent

The total angular momentum of the electromagnetic fields can be represented in two common forms: one in terms of the electric field and the transverse magnetic vector potential $\bm{A}^{\perp}$, and the other in terms of the magnetic field and the transverse electric vector potential $\bm{C}^{\perp}$. In this section, we show that the two formulations are not generally equivalent and that they describe distinct physical situations in which gauge invariance becomes essential. They coincide only in special limits, for example, for transverse fields, and in the monochromatic limit.

\subsection{Belinfante and canonical electromagnetic spin}\label{subsecc_spin_boundary}

%%%%%%%%%%%%%%%%%%%%%%%%%%%%%%%%%%%%%%%%%%%%%%%%%%%%%%%%%
%\textbf{Spin precession in vacuum}
%%%%%%%%%%%%%%%%%%%%%%%%%%%%%%%%%%%%%%%%%%%%%%%%%%%%%%%%%
The separation of electromagnetic angular momentum into spin and orbital components remains a fundamental topic in electrodynamics~\cite{Leader2014}. For radiation fields vanishing at infinity, the \textit{canonical angular momentum} decomposes cleanly into spin and orbital parts \cite{Stewart2005May,Leader2014}
\begin{align}
\bm{J}_{\text{can}} = 
\underbrace{\int \! d^3r \, \left(\mathbf{B} \times \mathbf{C}^\perp\right)}_{\text{spin}} 
+ \underbrace{\int \! d^3r \, B_i \left(\mathbf{r} \times \nabla C^\perp_i\right)}_{\text{orbital}},
\label{eq:canonical_angular_momentum}
\end{align}
where the classical \textit{orbital angular momentum} (OAM) density resembles the quantum OAM operator $\bm{L}_{\rm{op}} =  -i\hbar\ \bm{r} \times \nabla$ acting on $\bm{C}^\perp$ components \cite{Cohen1997}. Meanwhile, the gauge-invariant \textit{Belinfante angular momentum} reads
\begin{equation}
\bm{J}_{\text{Bel}} = \epsilon_0 \int \! d^3r \, \bm{r} \times \left( \bm{E} \times \bm{B} \right).
\label{eq:belinfante_angular_momentum}
\end{equation}
It has been noticed in several works that when the electromagnetic fields do not vanish at spatial infinity, these expressions differ by the following surface term \cite{Leader2014,Stewart2005May}
\begin{equation}
\label{s.t.}
    \int d^3r \, \nabla \cdot \bm{\mathsf{T}} = \oint_\Gamma d 
    \bm{\Gamma} \cdot \bm{B} \, (\bm{r} \times \bm{C}^\perp),
\end{equation}
where $\Gamma$ is a closed surface and $\bm{\mathsf{T}} = \bm{B} \otimes (\bm{r} \times \bm{C}^\perp)$.  

While deriving the expression for the spin angular momentum of non-monochromatic electromagnetic fields, several key aspects must be carefully considered. First, the Belinfante angular momentum represents a conserved and explicitly gauge-invariant quantity, and, in principle, only gauge-invariant observables can be physically measurable~\cite{Aharonov2015,Leader2016}. In contrast, the canonical spin density originates from the canonical angular-momentum tensor and is formulated in terms of the electromagnetic vector potential A, which is generally gauge dependent~\cite{Bliokh2017,Leader2014,Leader2016}. At the same time, however, the canonical spin angular momentum has been directly verified experimentally in optical measurements~\cite{antognozziDirectMeasurementsExtraordinary2016, liuThreeDimensionalMeasurementHelicityDependent2018, neugebauerMagneticElectricTransverse2018, shiStableOpticalLateral2022}. This apparent tension raises a fundamental question concerning how a quantity derived from a gauge-dependent formalism can manifest itself in experiments. This issue has been discussed in detail in the literature~\cite{Leader2016,Stewart2005May}.

The remainder of this section resolves this conceptual issue by clarifying the conditions under which different spin formulations become physically meaningful and by identifying a proper gauge-invariant expression for the spin density and total electromagnetic spin of non-monochromatic fields in the presence of sources.

\subsection{Gauge-invariant SAM of non-monochromatic fields with sources}
A vector field is said to be transverse (longitudinal) if and only if it is divergence-free (curl-free). In reciprocal space, a transverse (longitudinal) field is perpendicular (parallel) to the wavevector $\bm{k}$ for all $\bm{k}$~\cite{Cohen1997}. The canonical spin angular momentum (SAM) arises from the canonical angular-momentum tensor, which is defined by the gauge-dependent electromagnetic vector potential $\bm{A}$. The conventional approach, therefore, is to retain only the \emph{transverse} (i.e., divergence-free) and gauge-invariant component of this potential~\cite{Bliokh2017,Cameron2012,Barnett2010}.

For radiation fields vanishing at infinity, the total SAM of free transverse fields can be expressed in multiple equivalent forms \cite{Aiello2022,Stewart2005May} (Appendix \ref{Spin_density_Instantaneous_Fields}).  A symmetric canonical spin formulation which respects electric-magnetic democracy is~\cite{Barnett2010} 
\begin{equation}
    \bm{S} = \frac{1}{2} \int d^3 r \left[ \epsilon_0 \bm{E} \times \bm{A}^{\perp} + \bm{B} \times \bm{C}^{\perp} \right],
    \label{Spin_Symmtric}
\end{equation}
where $\bm{A}^{\perp}$ and $\bm{C}^{\perp}$ are the transverse parts of the electric and magnetic vector potentials, respectively. This expression for spin density is only valid for \emph{transverse electromagnetic fields} in the absence of both charge and current sources \cite{Aiello2022}. This symmetrization of the spin, however, becomes invalid for longitudinal fields and near-fields zone, since the transverse vector potentials are simultaneously gauge-invariant only for radiation fields \cite{Nienhuis2016}. Moreover, the presence of longitudinal components or sources breaks the duality between the electromagnetic fields.

A more general expression that is applicable to both \emph{transverse and longitudinal fields} in the presence of current source was found by Stewart based on Helmholtz's decomposition theorem (magnetically biased SAM) \cite{Stewart2005May}
\begin{equation}
   \bm{S}_{\textit{m}} =  \int d^3 \bm{r} \, \bm{B} \times \bm{C}^{\perp},
   \label{Spin_Stewart}
\end{equation}
where the electric vector potential is a non-local function of the magnetic field derivative. Here {\it magnetically} biased and analogously {\it electrically} biased refer to which field (magnetic or electric) is taken as the primary variable in defining the spin density. Crucially, Eqs. (\ref{Spin_Symmtric}) and (\ref{Spin_Stewart}) are derived under the assumption of vanishing fields at spatial infinity \cite{Stewart2005May}. Hence, the surface term is irrelevant. Notably, Eq.~\eqref{Spin_Stewart} offers the broadest validity, applying to instantaneous fields even in the presence of current sources (Appendix \ref{Spin_density_Instantaneous_Fields}). In the radiation zone, the electric and magnetic fields are dual-symmetric. However, the current source breaks electric-magnetic democracy in the near-field zone. 

An equally general expression can be obtained from the Helmholtz decomposition of the magnetic field. The electrically biased spin angular momentum of an instantaneous electromagnetic field in the presence of both transverse and longitudinal electric components is given by~\cite{Cohen1997}
\begin{equation}
   \bm{S}_{\textit{e}} =  \epsilon_0 \int d^3\bm{r} \, \bm{E} \times \bm{A}^{\perp},
   \label{Spin_Cohen}
\end{equation}
where both the electric and magnetic vector potentials are required to be transverse. This restriction, however, does not apply to the electromagnetic fields $\bm{B}, \bm{E}$ appearing in Eqs.~(\ref{Spin_Stewart})–(\ref{Spin_Cohen}).

The Belinfante total angular momentum density, given in Eq.~(\ref{eq:belinfante_angular_momentum}), is gauge-invariant because it does not involve the vector potentials. In contrast, neither the magnetically biased spin [Eq.~(\ref{Spin_Stewart})] nor the electrically biased spin [Eq.~(\ref{Spin_Cohen})] is gauge-invariant for \textit{arbitrary} electromagnetic configurations. Consequently, one must adopt the formulation that remains gauge-independent for the \textit{specific} electromagnetic problem under consideration, based on the two key properties of the examined field: transversality and electromagnetic duality, and include the surface term~[Eq.~\eqref{s.t.}] if the field does not vanish in such as a near-field. The gauge-invariant form corresponds to the physically measurable quantity, a realization that has also been used to reconcile differing interpretations of spin in particle and laser physics \cite{Leader2016}.

For example, let us consider a circularly polarized plane wave in a longitudinal static magnetic field $\bm{B} = \bm{B}^{\perp}_\text{circ} + \bm{B}^{\parallel}$. In this case, the field has a longitudinal magnetic field component and is not electromagnetically dual, hence the electrically biased formulation of spin is gauge-dependent and incorrectly predicts a nonzero orbital angular momentum (Appendix~\ref{Appendix:electrically-biased-formulation-constant-magnetic-field}). In contrast, the spin for same field configuration is gauge-independent in the magnetically biased formulation and exhibits pure spin angular momentum (SAM), in agreement with Eq.~(\ref{Spin_Stewart}) (Appendix~\ref{PrecessionBeff}). The electrically biased form, Eq.~\eqref{Spin_Cohen}, provides the proper gauge-invariant formulation and consistently separates spin and orbital contributions for the electromagnetically dual, non-monochromatic case of a circularly polarized wave in a longitudinal static electric field $\bm{E} = \bm{E}^{\perp}_\text{circ} + \bm{E}^{\parallel}$.

Similarly, the spin for superposition of a circularly polarized plane wave propagating along the $z$-axis and an orthogonally propagating plane wave with magnetic field linearly polarized along the $z$-axis is gauge-invariant only within the magnetically biased formulation (Appendix~\ref{App_Precessiing_Plane_Waves}). When the linear polarization of the second plane wave is switched to orthogonal one with $z$-oscillating electric field, the spin becomes gauge-invariant only in the electrically biased formulation. The selectivity of definitions here is connected to the breaking of magnetic-electric duality of such composite fields by construction, even though both constituent plane waves are transverse.

Finally, both the electrically and magnetically biased formulations are gauge-invariant for purely transverse electromagnetic fields. In this limit, where longitudinal components are absent, the spin can be expressed in a symmetric form due to the restoration of electromagnetic duality.

In most experimental configurations where the surface term vanishes, the Belinfante spin angular momentum reduces to the canonical spin angular momentum. This point is discussed in detail by Stewart~\cite{Stewart2005May}. Hence, the analysis presented in the next section on electromagnetic spin precession due to wave interference is independent of the distinction between the canonical and Belinfante spin angular momenta.

Gauge dependence couples the vectorial and spatial structure of electromagnetic fields \cite{Van1993,Van1994,Cohen1997}. 
Fernandez \emph{et al.} demonstrated that, except for a single plane wave, the field’s total angular momentum (AM) remains decoupled from its vectorial degrees of freedom even within the paraxial regime \cite{Fernandez2014}. 
For any field exhibiting spin precession, SAM must be gauge-invariant, as required by Eqs.~(\ref{Spin_Stewart}--\ref{Spin_Cohen}), and must satisfy the transversality condition $\mathbf{k}\!\cdot\!\mathbf{F}=0$ point-wise in Fourier space where $\bm{F}$ is the electric or magnetic field in reciprocal space \cite{Cohen1997,Van1994}. 
These conditions ensure that the angular momentum is purely spin in origin \cite{Li2009}. 
Consequently, the electromagnetic configurations considered here possess total angular momentum that reduces entirely to SAM, with no orbital contribution; see Appendices~\ref{orbital_angular_momentum_Bc_plus_B0} and \ref{Appendix:Orbital_angular_momentum}. The electromagnetic configurations that produce spin precession in the next sections corresponding to polarization interference without interference in the intensity \cite{Shevchenko2019} and pure SAM.

\subsection{Gauge-invariant SAM density of non-monochromatic fields with sources} \label{subsec_spin_density_nonmonochromatic_fields}

It is worth mentioning that the spin density behaves differently from spin. The spin density for transverse monochromatic electromagnetic fields is well-studied~\cite{Bliokh2017}. It can be expressed in terms of its magnetic-biased form $\mathbf{s}_{\text{m}} = \tilde{\mathbf{B}} \times \tilde{\mathbf{C}}^\perp$ or its electric-biased form $\mathbf{s}_{\text{e}} = \epsilon_0 \tilde{\mathbf{E}} \times \tilde{\mathbf{A}}^\perp$, or as a combination of its electric and magnetic parts $\mathbf{s} = \frac{1}{2} (\bm{s}_{\text{m}} + \bm{s}_{\text{e}})$ in case of free transverse electromagnetic fields~\cite{Bliokh2025,Bliokh2017,Bliokh2014,Alexeyev1999}. 
All quantities marked with tilda ($\tilde{\psi}$) denote the complex amplitudes of their respective fields or potentials, where the full physical quantities are given by $    \psi(\mathbf{r}, t) = \operatorname{Re}\left[\tilde{\psi}(\mathbf{r})e^{-i\omega t}\right]$. 
Importantly,  the electrically- and magnetically-biased expressions for spin densities are equal for time-harmonic (monochromatic) fields~\cite{Bliokh2025}.

The instantaneous spin density for non-monochromatic electromagnetic fields, on the other hand, remains poorly understood. While electromagnetic spin density is a local quantity, its description requires nonlocal potentials~\cite{Bialynicki2014,Stewart2005May}
\begin{equation}
    \begin{split}
        \bm{C}^\perp & = - \int d^3\bm{r}' \frac{\partial_t \bm{B}(\bm{r}',t)}{4\pi|\bm{r}-\bm{r}'|} ,\\
        \bm{A}^\perp &= \nabla \times \int d^3\bm{r}' \frac{\bm{B}(\bm{r}',t)}{4\pi|\bm{r}-\bm{r}'|}, 
    \end{split}
\end{equation}
making both spin density sensitive to \emph{global field properties and boundaries}. 

Unlike the total spin, the spatial divergence term plays an essential role in spin-density calculations and cannot be neglected. The magnetically biased total angular momentum density is \cite{Stewart2005May, Leader2014} 
\begin{equation}
    \bm{j}_{\text{m}} = 
    \bm{B} \times \bm{C}^{\perp} 
    - (\bm{B} \cdot \nabla)(\bm{r} \times \bm{C}^{\perp}) 
    + B_i \left(\bm{r} \times \nabla C^\perp_i\right),
    \label{eq:magnetic_spin_density}
\end{equation}
where $\bm{B} \times \bm{C}^{\perp}$ denotes SAM density, and $B_i (\bm{r} \times \nabla C^\perp_i)$ the orbital angular momentum (OAM) density. The second term,  
$(\bm{B} \cdot \nabla)(\bm{r}  \times \bm{C}^{\perp}) = \nabla \cdot [\bm{B} \otimes (\bm{r} \times \bm{C}^{\perp})] $,  
is often disregarded because of its explicit position dependence. However, it is indispensable for local angular momentum conservation near the current source and in unbounded electromagnetic field configurations such as plane waves \cite{Stewart2005May,Leader2014}. Its physical character whether spin-like, orbital, or mixed-depends on the fields configuration and current source nature. In many magnetically biased and non-monochromatic systems, this term represents genuine SAM contribution, including in the superposition of orthogonally propagating circularly and linearly polarized plane waves of different frequencies, and in the near-field spin density generated by precessing magnetic dipoles (see Appendix~\ref{PrecessionBeff}--\ref{Appendix_Magnetically-biased_spin_density}).

%\textcolor{green}{??? also add how do we know os contribution is orbital-like or spin-like}

The spatial divergence term exhibits spin-like character in an electromagnetic system where its functional form reduces to the canonical expression $\bm{B} \times \bm{C}^\perp$ as it occurs for transverse plane waves (Appendix~\ref{Appendix_superpositon_transverse_plane_waves}) and precessing magnetic dipoles (Appendix~\ref{Appendix_Magnetically-biased_spin_density}). Such terms must %always 
correspond to local field rotation (polarization changes)~\cite{Van1993}. In all our studied configurations (with vanishing orbital angular momentum density; Supplementary Section~\ref{Appendix:Orbital_angular_momentum}), the simplified Belinfante spin density reads 
\begin{equation}
    \bm{s}_{\text{m}} = \bm{B} \times \bm{C}^{\perp} 
    - (\bm{B} \cdot \nabla)(\bm{r} \times \bm{C}^{\perp}), 
    \label{eq:reduced_spin_density}
\end{equation}
with both terms have spin-like behavior. 

Invariance of spin density under translation provides another support for the validity of Eq.(\ref{eq:reduced_spin_density}) as a spin density equation. As an example, considering the electromagnetic spin and spin density emitted by a dipole, the spin has no position dependence and is invariant under translation [Eq.~\eqref{eq:Spin_dipole}]. Meanwhile, the spin density is functionally invariant through its Dirac-delta function associated with the source contact term [Eq.~\eqref{eq:spin_density_dipole}]. 

The Belinfante's spin density in Eq. (\ref{eq:reduced_spin_density}) is also consistent with the continuity equation [Eq.~\eqref{eq:spin_continuity_with_source}] where the right-hand-side torque density $\bm{M} \times \bm{B}$ emerges from magnetization coupling to light spin. It correctly reduces to the time harmonic and free space limits. As before, Belinfante's SAM coincides with the canonical magnetically-biased SAM when the surface term vanishes (Appendix~\ref{appendix:total_spin}).

%\textcolor{blue}{ Belinfante's SAM reduces to the corresponding canonical SAM as the surface term is not relevant experimentally.}

%The electrically-biased total angular momentum density \cite{Cohen1997,Leader2014},
%\begin{equation}
%\begin{split}
%\bm{j}_{\mathrm{e}} 
%  &= \epsilon_0 \Big[ 
%      \bm{E} \times \bm{A}^\perp 
%      - (\bm{E} \cdot \nabla)(\bm{r} \times \bm{A}^\perp) \\
%  &\quad\;\;
%      + E_i \bigl(\bm{r} \times \nabla A^\perp_i\bigr)
%    \Big],
%\end{split}
%\label{eq:electric_angular_momentum_density}
%\end{equation}
%is commonly used to separate spin and orbital components in optical and laser physics. 
%A complete derivation is provided in Appendix~\ref{spin_precession_dipole}.

\subsection{Electromagnetic spin dynamics with sources}\label{subsec_spin_density_sources_continuity_equation}
While the Landau-Lifshitz equation is widely used for describing magnetization dynamics in ferromagnetic materials, its extension to the dynamics of optical spin represents unexplored physics \cite{Landau1992,Zvezdin2022}. As we show below, a new source term in the continuity equation of the electromagnetic spin couples the photonic and material spin degrees of freedom, revealing the nature of such light-matter interaction mechanism. 
%This formulation provides a unified description of coupled spin dynamics across electromagnetic and condensed matter systems.

The conventional continuity equation for spin density, $\partial_t s_i + \partial_j \Sigma_{ij} = 0$, holds only for monochromatic source-free fields. For systems with current sources, 
using definition~\eqref{eq:reduced_spin_density} for spin density, taking its time derivative, and utilizing Maxwell equations with sources, we derive a generalized form for the continuity equation, (see Supplementary materials for complete derivation in Section~\ref{Appendix_continuity_equation}) 
\begin{equation}
    \partial_t \bm{s}_\text{m} + \nabla \cdot \left(\bm{\mathsf{\Sigma}} + \bm{\mathsf{\Phi}}\right) = -\bm{M} \times \bm{B},
    \label{eq:spin_continuity_with_source}
\end{equation}
The source term $-\bm{M} \times \bm{B}$ arises exclusively from the $\bm{B} \times \bm{C}^\perp$ component of the spin density, reflecting the \textit{spin torque} density produced by the magnetization. This term vanishes in source-free regimes but becomes crucial in electromagnetic systems with magnetic sources. %The time averaged $ \langle \bm{\mathsf{\Sigma}} + \partial_t \bm{\mathsf{T}}\rangle$ reduces to the known expression for spin current density in the time harmonic limit (Appendix \ref{continuity_equation_with_sources}). 

The total spin current density consists of two parts: $\bm{\mathsf{\Sigma}}$ arises from $\bm{B} \times \bm{C}^\perp$ and the term $\bm{\mathsf{\Phi}} = \partial_t \bm{\mathsf{T}}$ is a direct consequence of $\nabla \cdot \bm{\mathsf{T}} = (\bm{B} \cdot \bm{\nabla})(\bm{r} \times \bm{C}^\perp)$. The $\bm{B} \times \bm{C}^\perp$ term alone reproduces the spin density in the harmonic limit as period average  $\langle \bm{\mathsf{\Phi}} \rangle =0$, but the full expression $\bm{s}_\text{m}$ is essential for consistency with local angular momentum conservation. 

%\blue{???needs link to SM general proof, or a proof for which examples it works, or we are saying about integral quantities (total AM) which are obviously gauge invariant (?)} \textcolor{red}{I think that the magnetically biased spin density is gauge invariant becasue the electtric vector potential is always transverse vector in case of no charge sources}. 
%
In magnetic systems, where the charge density $\rho = 0$ and current density $\bm{j} \neq 0$, the electric vector potential $\bm{C}^\perp$ is purely transverse ($\nabla \cdot \bm{C}^\perp =0$)  according to Helmholtz's decomposition theorem \cite{Stewart2005May}.
The dependence of spin density on the nonlocal vector potentials (both electric $\bm{A}^\perp $and magnetic $\bm{C}^\perp$) naturally introduces boundary-sensitive terms in the continuity equation. This nonlocality, arising from the Helmholtz decomposition of the fields, explicitly couples the spin dynamics to the system boundaries~-- a fundamental consequence of the gauge-invariant formulation~\cite{Bialynicki2014}. 
Our choice of spin density definition %theoretical framework 
unifies spin dynamics in both source-free and magnetically-biased systems, enabling accurate modeling and gauge independent formulation of photonic spin precession. %spin-transfer phenomena.

%\blue{Add volume-integrated result: very important!!!} 

%\textcolor{blue}{In summary, the choice between magnetically and electrically biased formulations should be guided by gauge invariance and by the continuity relations for the SAM density. The magnetically biased Belinfante and canonical SAM provide the natural description for non-monochromatic fields generated by magnetic sources, such as oscillating or precessing magnetic dipoles. In these cases, the electric vector potential is uniquely determined and gauge-invariant for purely current-driven fields containing both transverse and longitudinal magnetic components. Conversely, the electrically biased Belinfante and canonical SAM formulations offer the appropriate gauge-invariant description for systems dominated by electric sources, such as time-varying electric dipoles or electromagnetic fields with longitudinal electric components.
%} 

\section{Electromagnetic spin precession}\label{sec_wave_inference_nosources}

It can be shown that monochromatic fields doe not precess (Appendix~\ref{Spin_Harmonic_waves}). In this section, we consider several instances of non-monochromatic electromagnetic fields, precession of spin density or total electromagnetic spin, and propose their possible realizations. 

\subsection{Spin precession resulting from rotating observer}
%\noindent

We start from the trivial and intuitively understandable case of photonic spin precession appearing in the rotating frame, 
%
%When light is viewed from a rotating frame, 
where even simple optical fields can acquire subtle frequency and phase shifts. Such effects are familiar from the Sagnac interferometer, where rotation produces a measurable fringe shift, and from the angular Doppler effect, where the frequency of circularly polarized light is shifted by the observer’s rotation \cite{Post1967,Lavery2013,Garetz1979,Garetz1981}. These phenomena reflect the coupling between electromagnetic angular momentum and mechanical rotation \cite{Speirits2014,Padgett2006}. While Maxwell’s equations are time-reversal symmetric in inertial frames, rotation introduces non-inertial terms that break this symmetry \cite{Coisson1973}. In a rotating frame, the electromagnetic spin obeys
\begin{equation}
\left( \frac{d \mathbf{S}}{dt} \right)_{\mathrm{rot}} = \mathbf{S}_{\mathrm{rot}} \times \boldsymbol{\Omega},
\label{precessionResult}
\end{equation}
where $\boldsymbol{\Omega}$ is the rotation velocity. This purely kinematic precession is the optical analogue of Larmor precession \cite{Bliokh200Coriolis}. Appendix~\ref{precessionrotating} derives Eq.~\eqref{precessionResult} from first principles, showing that it arises from rotation-induced charge and current densities in Maxwell’s equations that break time- and rotation-reversal symmetry. The precession frequency $\Omega$ produces an angular Doppler shift $\sigma\Omega$ ($\sigma=\pm1$ for right- and left-circular polarization), completing the mechanical-electromagnetic analogy to Coriolis effects~\cite{Speirits2014,Garetz1981,Garetz1979,Lavery2013}.

%\textcolor{green}{Abanub, please briefly describe this case and give a reference to the corresponding Supplementary section} 

%\blue{maybe here connection to optical gyrocscopes? rotating observer -> frequency shift } 

\subsection{Circularly polarized plane wave in longitudinal external magnetic field }\label{subsec_circualrly_polarized_wave_constant_field}

%\blue{maybe this section could be merged a bit more with the next one? since it's just a special case}

\begin{figure*}[t!]
\includegraphics[width = 0.96 \textwidth]{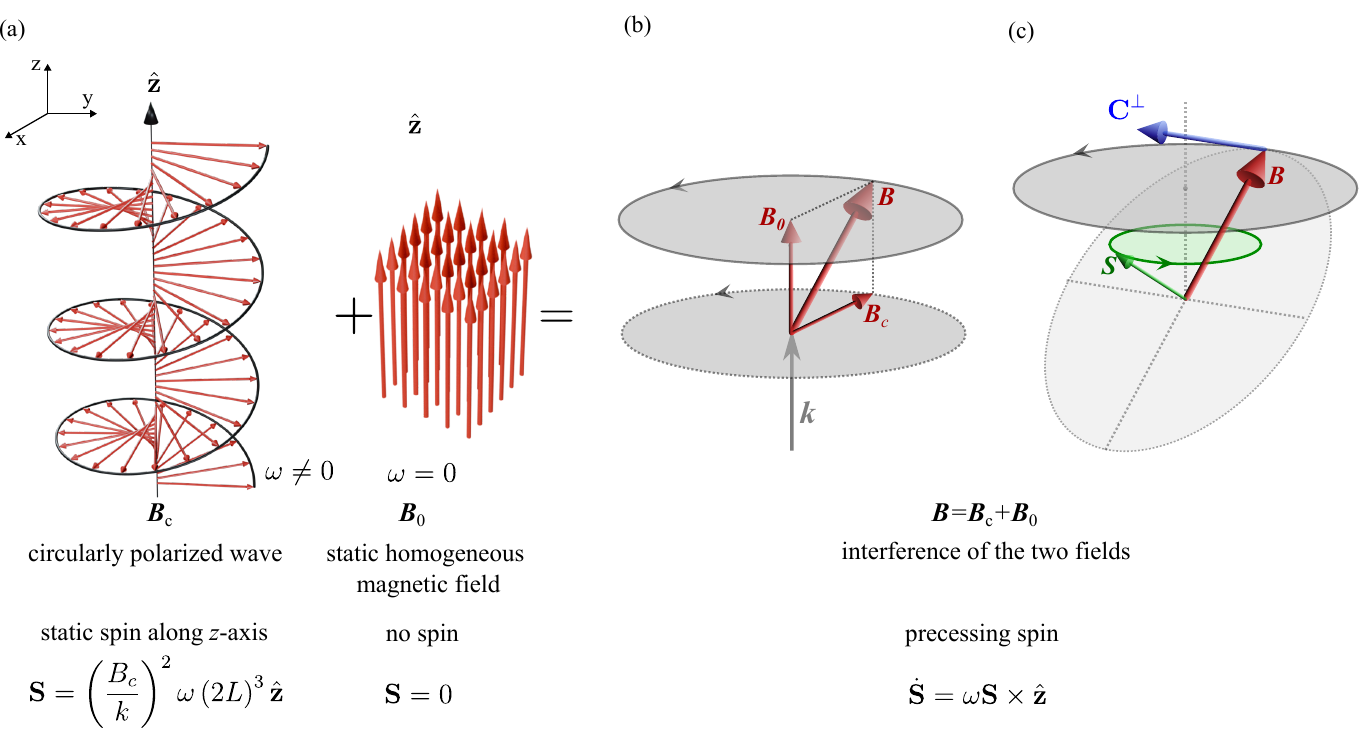}
\caption{
(a) A circularly polarized electromagnetic wave of frequency $\omega$, carrying a static spin angular momentum oriented along the $z$-axis, interacts with a homogeneous static magnetic field $\bm{B}_0$. (b) Their superposition generates a total magnetic field $\bm{B} = \bm{B}_c + \bm{B}_0$ that undergoes precession. (c) The instantaneous spin of the resulting electromagnetic field precesses at the driving frequency $\omega$, remaining perpendicular to the plane defined by the static magnetic field $\bm{B}_0$ and the transverse vector potential $\bm{C}^\perp$.
}% and precesses at the same frequency. 
\label{fig:precession_constant_B0}
\end{figure*}

Inspired by the structure of the magnetic field in the ferromagnetic resonance, next we analyze the electromagnetic spin dynamics resulting from the interference of a circularly polarized wave
\begin{equation*}
    \bm{B}_\text{c}(z,t) = B_\text{c}\big[\cos(k z - \omega t )\hat{\bm{x}} + \sin(kz-\omega t )\hat{\bm{y}}\big]
\end{equation*}
with a static magnetic field \(\bm{B}_0\), see Fig.~\ref{fig:precession_constant_B0}. The spatial profile of the wave and the static field vanishes at the boundary (i.e., measurement device), thereby breaking both translational and rotational symmetry. Furthermore, since the fields in a realistic setup vanish at the boundary of a large volume~\cite{Stewart2005May}, the surface term $\oint_\Gamma d 
    \bm{\Gamma} \cdot \bm{B} \, (\bm{r} \times \bm{C}^\perp) = 0$. The orbital angular momentum contribution, \(\int d^3r \, B_i (\bm{r} \times \nabla C^\perp_i)\) is also zero (Appendix \ref{PrecessionBeff}). Consequently, the Belinfante angular momentum reduces to the canonical magnetically-biased SAM:
\begin{equation}
    \bm{J}_{\text{Bel}} = \bm{S}_{\text{m}} 
    = \int d^3 r\, \bm{B} \times \bm{C},
    \label{eq:total_spin_interference}
\end{equation}
While the circularly polarized wave alone carries constant spin along $\hat{\bm{z}}$ direction, the interaction with the time-independent field induces precession described by (Appendix \ref{PrecessionBeff}):

\begin{equation}
\bm{S} = 
a \cos\omega t \hat{\bm{x}}
+ a \sin\omega t \hat{\bm{y}}
+ b \hat{\bm{z}},
\label{eq:spin_precession}
\end{equation}
where  $a =\epsilon_0\left(8 B_c B_0 /k^3 \right) \sin(k L) L^2$ and $b =\epsilon_0 \left(B_c/k\right)^2 \omega L^3 $ with $\left(2L\right)^3$ being a normalization volume. 
%\textcolor{green}{write this expression more compactly} 
The transverse components ($S_x$, $S_y$) exhibit $\omega$-periodic precession, while $S_z$ remains constant, matching the intrinsic spin of the circularly polarized wave~\cite{Stewart2005May}. The spin $\bm{S}$ in Eq.~\eqref{eq:spin_precession} obeys a Landau-Lifshitz-like %-Gilbert 
 equation, 
\begin{equation}
    \frac{d \bm{S}}{dt} = \omega \bm{S} \times \hat{\bm{z}},
    \label{LLG-like-equation}
\end{equation}
demonstrating the precessional dynamics induced by the electromagnetic fields.

Two approaches are possible to model a finite constant $z$-directed magnetic field: 
(i) a transversally finite wave (propagating along $x$ with $z$-polarized $\bm{B}$ field) in the static limit $\omega_2 \to 0$ as shown in the example blow, or 
(ii) an external $z$-confined static magnetic field.

The circular polarization's spin maintains fixed orientation under time reversal (handedness change), but the applied $\bm{B}_0$ breaks time-reversal and inversion symmetry, enabling precession. 
\textcolor{red}{} 
This manifests in the sinusoidal oscillations of $S_x$ and $S_y$, with amplitude modulated by the spatial interference factor $\sin(kL)/k$. The precession frequency $\omega$ and interaction length $L$ together govern the spin dynamics: $\omega$ sets the temporal scale of precession, while $L$ determines the spatial region where field interference occurs. Their product $kL = (2\pi/\lambda)L$ in the amplitude factor $\sin(kL)/k$ reflects how the wave's phase variation across the interaction region modulates the precession strength.

Figure~\ref{fig:precession_constant_B0}(a) illustrates the precessional motion arising from the interference of a circularly polarized wave with a static magnetic field. The total magnetic field \(\bm{B}\) precesses around the external magnetic field \(\bm{B}_0\) as shown schematically in Fig.~\ref{fig:precession_constant_B0}(b). The electric vector potential \(\bm{C}^\perp\) remains tangential to the precessional trajectory traced by \(\bm{B}\). The spin \(\bm{S}\) likewise undergoes precession in a plane orthogonal to the one spanned by \(\bm{B}\) and \(\bm{C}^\perp\). 

\begin{figure*}[t!]
\includegraphics[width = 0.99 \textwidth]{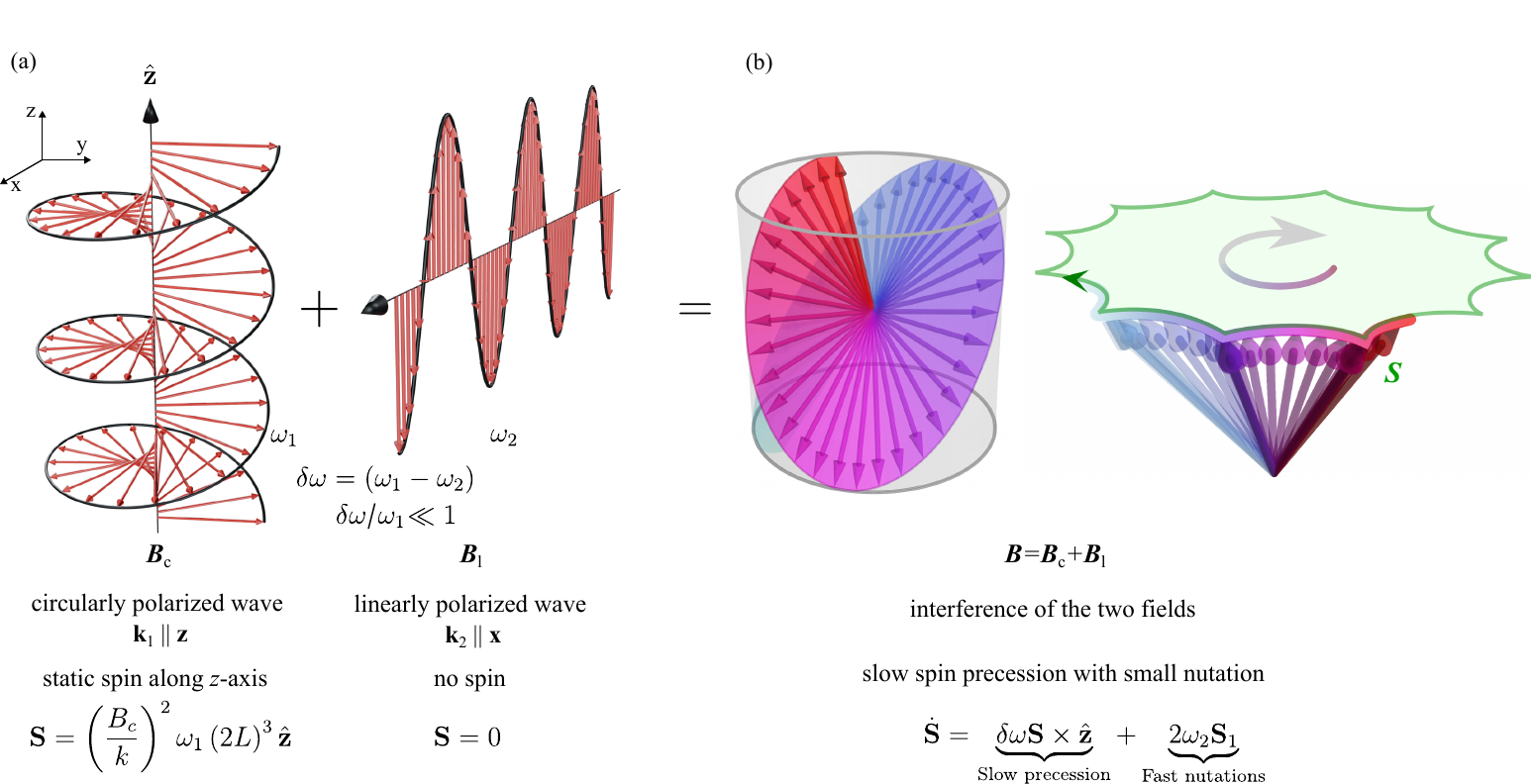}
\caption{
%illustrate precessional motion resulting from mixing of circularly and linearly polarized plane waves propagating in orthogonal directions with frequencies $\omega_{1,2}$. 
(a) A circularly polarized wave at frequency $\omega_1$ propagating along $z$-axis (with static spin along the $z$-axis) interfering with a linearly $z$-polarized wave at frequency $\omega_2$ propagating along the $x$-axis. 
(b) Total magnetic field trajectory of the combination evolves along quasi-ellipsoidal trajectories gradually rotating at the beating frequency $\delta \omega \equiv \omega_1 - \omega_2$. %for of circularly and linearly polarized plane waves propagating in orthogonal directions with distinct frequencies $\omega_{1,2}$. 
(b) The instantaneous Belinfante electromagnetic spin experiences a slow precessional motion at the difference frequency $\delta \omega$ with nutations of small amplitude %$\propto \delta \omega / \omega_{1,2}$ 
at the sum frequency $\omega_1 + \omega_2$ in the quasi-monochromatic limit $\delta \omega \ll \omega_{1,2} $.
Parameters: $\omega_2 = 0.2 \omega_1$, $B_\text{c} = B_\text{l}$. 
}
%\label{fig:precession_wave_mixing}
\label{fig:nutation}
\end{figure*}

\subsection{Orthogonal plane-wave interference~\label{2PlaneWaves}}\label{subsec_two_wave_interference}

%\blue{intersecting beams!}

Experimentally, however, it could be more difficult to generate and control constant fields. %Interference between plane waves is more common and is known to lead to many interesting phenomena \cite{Bliokh2025}. 
In this endeavor, we consider the interference between a circularly polarized wave and a linearly polarized wave propagating at orthogonal directions, such that the magnetic field of the linearly polarized one is along the propagation direction of the circularly polarized one, see Fig.~\ref{fig:nutation}. As before, the Belinfante angular momentum is equal to the magnetically-biased spin $\bm{J}_{\rm{Bel}} = \bm{S}_\text{m}$ (below we omit the subscript m %in Eq.(\ref{eq:total_spin_interference}) 
for clarity)
when the fields vanish at the boundary. This wave interference generates a time-dependent spin angular momentum $\bm{S}$ that decomposes into two distinct precessing components, $\bm{S}_1$ and $\bm{S}_2$, each evolving at characteristic frequencies (see Appendix \ref{App_Precessiing_Plane_Waves}) 
\begin{align}
    \frac{d\bm{S}_1}{dt} &= (\omega_1 + \omega_2) \bm{S}_1 \times \hat{\bm{z}}, \label{S1dot} \\[10pt]
    \frac{d\bm{S}_2}{dt} &= (\omega_1 - \omega_2) \bm{S}_2 \times \hat{\bm{z}}. \label{S2dot}
\end{align}
In the quasi-monochromatic limit $\delta \omega/(\omega_1 + \omega_2) = (\omega_1 - \omega_2)/(\omega_1 + \omega_2) \ll 1$, $S_1 / S_2 \sim \delta \omega / \omega_{1,2} \ll 1$, and the spin dynamics follow  
\begin{equation}
    \frac{d \bm{S}}{dt} = \underbrace{\delta \omega \, \bm{S} \times \hat{\bm{z}}}_{\text{slow precession}} + \underbrace{2 \omega_2 \bm{S}_1 \times \hat{\bm{z}}}_{\text{fast nutations}},
\end{equation}
The first term represents the slow precession of the total SAM while the second term represents the fast nutations. 

Figure~\ref{fig:nutation}(a) illustrates the interaction between a circularly polarized wave and a linearly polarized wave, while Fig.~\ref{fig:nutation}(b) depicts the trajectory of the total magnetic field for circularly and linearly polarized plane waves propagating orthogonally with distinct frequencies $\omega_{1,2}$. Figure.~\ref{fig:nutation}(c) schematically shows dynamics of Belinfante electromagnetic spin, where it undergoes precession at the difference frequency $\delta \omega \equiv \omega_2 - \omega_1$ accompanied by small-amplitude nutations ($\sim B_\text{c} \delta \omega / \omega_{1,2}$) at the sum frequency $\omega_1 + \omega_2$ in the quasi-monochromatic limit ($\delta \omega \ll \omega_{1,2}$).
%\blue{something about nearly dendegerate waves -- or equivalently quazi-monochromatic waves}
On the timescales of $\tau$ relevant for the potential test particle such that $1/\omega_{1,2} \ll \tau \ll 1/\delta\omega$, the photonic spin will experience pure precession while small nutations will average out.

It follows from Eqs.~\eqref{S1dot}--\eqref{S2dot}, we note that when \(\omega_1 = \omega_2 = \omega\), the spin dynamics yields \(\dot{\bm{S}} = 0\), as expected for monochromatic fields (Appendix~\ref{Spin_Harmonic_waves}). In the limit \(\omega_2 \rightarrow 0\), the system correctly reduces to the case of a circularly polarized wave interfering with a static magnetic field.
%\textcolor{red}{Max, MAYBE IT IS IMPORTANT: IN A HOMOGENOUS MEDIUM ELECTROMAGNETIC SPIN PRECESSION COULD ARISE FROM A CIRCULARLY POLARIZED WAVE INTERFERING WITH A STANDING WAVE}

\subsection{Near-field of a precessing magnetic dipole} \label{sec_precessing_magnetic_dipole}

Looking at the ferromagnetic resonance from the opposite perspective, we also consider the field generated by the precessing %point 
magnetic dipole. 
We find that in the near-field, the electromagnetic spin density also locally precesses synchronously, with an off-axis maximum of spin amplitude rotating with the frequency of magnetic dipole precession.

The precession of electromagnetic spin reveals a fundamental analogy with well-known magnetic phenomena. %Remarkably, 
The interference between a circularly polarized wave and a static magnetic field~-- precisely the configuration producing electromagnetic spin precession in Eq.~(\ref{eq:spin_precession})~-- mirrors the conditions for ferromagnetic resonance (FMR) in materials. This correspondence is not accidental as both phenomena emerge from the universal dynamics of angular momentum under magnetic bias. While FMR describes %electron 
spin precession in ferromagnets~\cite{Gurevich2020}, the electromagnetic counterpart demonstrates identical precessional behavior for photon spin. This parallel dynamics establish a profound link between optical and magnetic systems, suggesting that photonic spin manipulation could emulate established spintronic phenomena in novel device architectures.

\begin{figure*}
\includegraphics[width = 0.99 \textwidth]{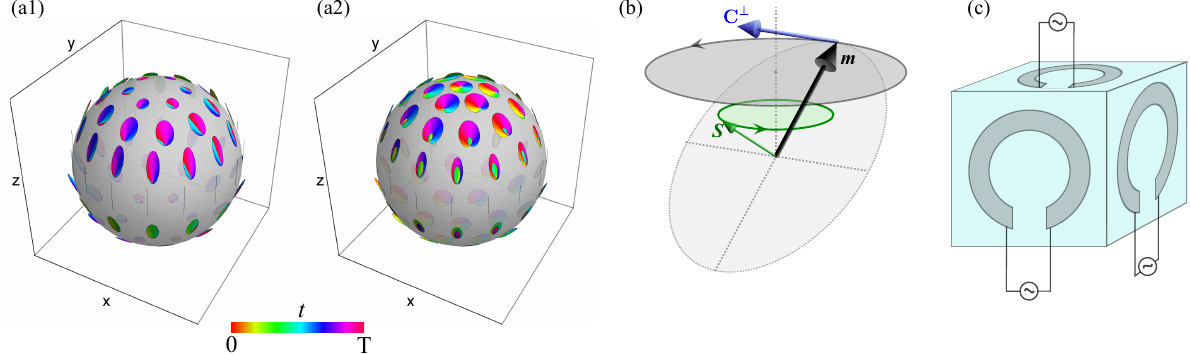}
\caption{
(a1-a2) Vector tips trajectories of local near-field electromagnetic spin density on a sphere of unit radius around the time-dependent point magnetic dipole (with time color-coded), for the cases of rotating dipole with $b_0 = 0$ (a1) and precessing dipole with $b_0 = 0.5 m_0$ (a2). 
(b) The Belinfante electromagnetic total spin experiences a synchronous precessional motion with the same frequency. 
(c) Possible realization of the precessing electromagnetic dipole as a split-ring meta-atom with individual external current drives for each split-ring, modeling three %corresponding 
time-dependent magnetic dipole components. }
\label{fig:spin_precession}
\end{figure*}

This connection between electromagnetic (EM) spin and condensed matter spin dynamics motivates our investigation of EM spin %emission 
of 
from precessing magnetic dipoles, see Fig.~\ref{fig:spin_precession}(b). 
%
%Note that besides natural magnetic materials, photonic precessing magnetic dipole could be emulated by the meta-atom with independently controlled split-ring elements, as shown schematically in Fig.~\ref{fig:spin_precession}(b). 
%
For a precessing dipole which carries purely spin angular momentum, the orbital angular momentum of the field is zero, see Appendix~\ref{Appendix:Orbital_angular_momentum}. The Belinfante magnetically-biased spin density reads (see Appendix \ref{Appendix_Magnetically-biased_spin_density})
\begin{widetext}
\begin{equation}
\begin{split}
\bm{s}_{\rm{m}} & = 2 \epsilon_0 \left(\frac{\mu_0}{4 \pi} \right)^2 \left[ \frac{(\hat{\bm{r}} \times \dot{\bm{m}}) (\hat{\bm{r}} \cdot \bm{m})}{r^4}  + \frac{4 \pi}{3} \, \frac{(\hat{\bm{r}} \times \dot{\bm{m}}) (\hat{\bm{r}} \cdot \bm{m})}{r} \, \delta\left( \bm{r} \right) \right]\\ 
& = 2 \epsilon_0 \left(\frac{\mu_0}{4 \pi} \right)^2 \left[ \frac{\, m_0 \, \omega \, \mathfrak{\xi}(x,y,z,t)}{r^5} + \frac{4 \pi}{3} \,\frac{\, m_0 \, \omega \, \mathfrak{\xi}(x,y,z,t)}{r^2} \delta(\bm{r})\right] \left[\Re \left(z \left( \hat{\bm{x}} -i \hat{\bm{y}} \right) e^{i \omega t} \right)  + \Im\left( \left(y -i x \right) e^{i \omega t} \right) \hat{\bm{z}} \right]
\end{split}
\label{eq:spin_density_dipole}  % Single label for the entire equation
\end{equation}
\end{widetext}
%\blue{$\epsilon_0 \mu_0$ etc.~-- check and correct here and in the Supplementary}
where the space-time dependent amplitude is defined as $\mathfrak{\xi}(x,y,z,t)  = b_0 \frac{z}{r} + m_0 \frac{x}{r}  \cos(\omega t)+ m_0 \frac{y}{r} \sin(\omega t)$. The first term dominates the %non-radiative 
near-field zone ($\sim r^{-4}$) while the second represents a localized spin density at the dipole location. The magnetically-biased formulation naturally captures both contributions, including the boundary-sensitive $(\bm{B}\cdot\nabla)(\bm{r}\times\bm{C}^\perp)$ term that becomes non-negligible near sources (Appendix \ref{Appendix:magnetically_biased_spin_dipole}). 
%The EM spin density is functionally invariant under translation due to the Dirac-$\delta$ function: The transformation $\bm{r} \rightarrow \bm{r} + \bm{d}$ would merely translate the spin source location from $\bm{r}_0 = \bm{0}$ to $\bm{r}_0 = \bm{d}$.

The %full 
local instantaneous spin density contains frequency components at $0$, $\omega$ and $2\omega$. Terms that are proportional to $m_0^2$ contain both a time-independent (DC) contribution and a $2\omega$ contribution, whereas terms proportional to $m_0 b_0$ oscillate at the fundamental frequency $\omega$ (see full derivation in Appendix \ref{appendix:frequency_decomposition}). 
%Interestingly, even in the monochromatic case when the dipole rotates in the $x$-$y$ plane with zero magnetic dipole moment along $z$-axis ($b_0 =0$), the spin density has two frequency components at $0$ and $2\omega$, which is connected with the presence of the source in the form of the magnetic dipole. 

Figures~\ref{fig:spin_precession}(a1) and \ref{fig:spin_precession}(a2) illustrate the precession of the spin density components at the unit sphere for two representative cases~-- rotating dipole ($b_0 = 0$) and precessing dipole ($b_0 = 0.5 m_0$). %in the two-dimensional cross-section at the $z = 0.5$ plane, as well as on the unit sphere. 
Compared to the rotating dipole [Fig.~\ref{fig:spin_precession}(a1)], spin density for the precessing dipole [Fig.~\ref{fig:spin_precession}(a2)] contains three frequency components and evolves along non-elliptic trajectories, with nonzero (rotating) spin density along the $z$ axis. 
Furthermore, as expected in the near field, the spin density pattern follows the magnetic dipole rotation as a solid structure (see Appendix~\ref{appendix:spin_density}). %The $x$- and $y$-components are represented by arrows. %, while the $z$-component is color-coded. 

On the other hand, the total EM spin in the near-field zone reads~(see Fig.~\ref{fig:spin_precession}(b) and Appendix~\ref{appendix:total_spin}) 
\begin{equation}
\begin{split}
\bm{S} & = \frac{4}{3} \epsilon_0\left(\frac{\mu_0}{4\pi}\right)^2 \frac{\bm{m} \times \dot{\bm{m}}}{a_0} \\
       & = \frac{4}{3} \epsilon_0 \left(\frac{\mu_0}{4\pi}\right)^2 \frac{m_0 \omega}{a_0}  \,    \Re \left[ b_0 e^{i\omega t} (\hat{x} - i \hat{y})
    - m_0 \hat{z} \right]
    \label{eq:Spin_dipole}
\end{split}
\end{equation}
where $a_0$ is a characteristic size of the magnetic dipole and $ \bm{m} \times \dot{\bm{m}}  = m_0 \omega  \,    \Re \left[ b_0 e^{i\omega t} (\hat{x} - i \hat{y})- m_0 \hat{z} \right]$. It is straightforward to show that $\bm{S}$ in Eq.~\eqref{eq:Spin_dipole} obeys the same Landau-Lifshitz equation~\eqref{LLG-like-equation} and precesses with the frequency $\omega$ as illustrated in Fig.~\ref{fig:spin_precession}(b).   %with different expression for spin. 
%The total spin in both cases also %, on the other hand, has only two frequency components $0$ and $\omega$ revealing pure precessional motion. 
%in the general case of a precessing dipole with nonzero z-component ($b_0 \neq 0$). 

%Our 
The derivation of the spin angular momentum continuity equation [Eq.~\eqref{eq:spin_continuity_with_source}] is %fully 
general and accounts for spin sources represented by the Dirac delta function term in Eq.~\eqref{eq:spin_density_dipole}. To complete the physical picture of electromagnetic spin precession, we now examine the associated spin current.
 The time-averaged spin current density, $\langle \bm{\Sigma} + \partial_t \bm{T} \rangle = \langle \bm{\Sigma} \rangle$ as $\langle \partial_t \bm{\mathsf{T}} \rangle = 0$, from a precessing magnetic dipole reveals intricate %fundamental 
 aspects of angular momentum transport:
\begin{equation}
\langle \bm{\mathsf{\Sigma}} \rangle = \frac{\epsilon_0}{2}\left(\frac{\mu_0}{4\pi}\right)^2 m_0^2\omega^2
\begin{bmatrix}
0 & 0 & -y/r^4 \\
0 & 0 & x/r^4 \\
-y/r^4 & x/r^4 & 0
\end{bmatrix}
\label{eq:spin_current_tensor}
\end{equation}
\\
%\textcolor{red}{What does symmetric structure of the spin current density tensor imply? } 
Vanishing diagonal components of the spin current density tensor %(zero diagonal elements) 
indicate that no net spin angular momentum is transported parallel to its own polarization axis in this system. Consequently, all spin currents in this system are purely transverse, flowing perpendicular to the spin polarization direction. The $\hat{\bm{x}}  \hat{\bm{y}} $ and $\hat{\bm{y}} \hat{ \bm{x}}$ terms represent in-plane spin transfer, while $\hat{\bm{x}} \hat{\bm{z}} $  and $\hat{\bm{y}} \hat{\bm{z}}$ components show out-of-plane spin transfer.

The $x/r^4$ and $y/r^4$ terms reveal a spatial pattern where spin angular momentum density circulates azimuthally around the dipole's precession axis ($z$-direction). This matches the expected symmetry for a precessing magnetic moment. The $1/r^4$ dependence signifies near-field dominance, contrasting with the $1/r^2$ decay of energy flux in dipole radiation. This confirms the spin current is primarily a non-radiative near-field effect. The rapid $1/r^4$ decay suggests that spin current mediation requires sub-wavelength dipole spacing, which is crucial for designing compact spin-wave devices.

\subsection{Possible Experiments}\label{sec_experiments}

\subsubsection{Quasi-monochromatic two-beam interference}

The two-wave interference setup proposed in Section~\ref{2PlaneWaves} holds promise in observing photonic spin dynamics (low-frequency precession with small nutations) in quasi-monochromatic fields when frequencies of the two waves are close, $|\omega_1 - \omega_2| \ll \omega_{1,2}$. %A realized in optical domain via interference of two orthogonal beams. 
%
%Polarization beating stands for a periodic variation of the field polarization state. Interfering an optical wave with its orthogonally polarized, frequency-shifted counterpart (e.g., its second harmonic) generates intricate polarization Lissajous curves rather than conventional elliptical polarization, enabling the creation of optical beams carrying fractional-order angular momenta \cite{Pisanty2019}. 
%
%To verify the predicted spin precession dynamics arising from the interference of orthogonally propagating circularly and linearly polarized waves with a small frequency offset, we propose a heterodyne interferometry experiment. 
A circularly polarized laser beam propagates along the $z$-axis, while a frequency-shifted copy (via an acousto-optic modulator, AOM) is linearly polarized along $z$ and directed along the $x$-axis using a polarizing beam splitter (PBS) \cite{Patra2005}. 
%Photonic spin of the resulting field could be detected via its action through the spin torque it exerts on the magnetic dipole particle placed in the interference region, manifesting in the magnetic moment mimicking the slow photonic spin precession dynamics. 
%\textcolor{red}{Indeed, (describe the magnetic moment dynamics) .........}

Anisotropic particles~\cite{Riccardi2023Feb,Toftul2024Oct}, such as optically levitated rotors driven by circularly polarized light~\cite{Zeng2024,Ahn2020,Xie2021}, have been shown to spin stably at MHz-GHz rates and to exhibit clear axis deviation and precession. The scattered light from the rotor’s orientation and rotation provides a direct readout of the electromagnetic spin-induced torque. Photonic spin precession could be similarly detected through the torque it exerts on optically levitated micro gyroscopes placed in the interference region of circularly and linearly polarized orthogonal beams. The circularly polarized light induces spin in the absorbing microscopic spheroid particle by transferring angular momentum from the light to the material. The orthogonal linearly polarized beam should cause %gyroscopic rotation with 
slow precession of the spinning spheroid~\cite{Zeng2024,Xie2021, Friese1998}. 

\subsubsection{Meta-atom: Split ring resonator}
%- microwave externally controllable 3D (split (?))-ring resonator as a meta-atom with precessing 3D magnetic moment. metasurface/ Article: how to excite split-ring designs 
%delay line (?) for each of the three groups of control voltages . Gneralize to chains or any other dynamics. 
%\begin{figure}[t!]
%  \centering
%  \includegraphics[width=0.65\columnwidth]{Fig13.pdf} % Adjust scale factor (0.5-0.9)
%  \caption{
%    Proposed experimental setup for a meta-atom
%  }
%  \label{fig:3}
%\end{figure}
%To emulate such physics, we propose an idea of a microwave experiment utilizing three-dimensional split-ring meta-atoms with currents in the split rings controlled individually via independent external AC or DC drives. 

To verify the predicted precession of electromagnetic spin in the near-field of the precessing magnetic dipole, we propose a microwave experiment using independently electrically controlled split-ring resonators (SRRs) emulating the %artificial 
precessing magnetic dipole components, see Fig.~\ref{fig:spin_precession}(c). 
A pair of orthogonally oriented SRRs, each loaded with a varactor diode, can be driven by phase-shifted AC signals (90$^\circ$ offset) to mimic the rotating transverse magnetic moment of a precessing dipole, while a third DC-driven SRR produces the static $z$-component. %, while a DC bias applied to the varactors sets the resonant frequency (analogous to the Larmor frequency). 
%The static $z$-directed moment could be emulated by embedding a ferrite disk within the SRR substrate, magnetized by an external DC magnetic field. 
%\textcolor{red}{Near-field scanning probes (e.g., a dipole antenna or magneto-optic sensor) could then map the spatial and temporal evolution of the electromagnetic spin density, with the AC drive frequency swept to match the theoretical precession dynamics.}
%Successful validation would establish SRRs as a versatile tool for emulating spin-dependent phenomena in classical electromagnetism. 
By constructing a chain from such SRR meta-atoms, the photonic analogue of spin waves can be excited.

This platform, being suitable to emulate general three-dimensional time dynamics for magnetic dipoles beyond precession, could lead to discovery of more unconventional photonic spin behavior, including in the far fields, opening new potential pathways for magnetic interactions and photonic spin transfer.

\section*{Conclusion}
Using a gauge-invariant formulation of electromagnetic spin angular momentum in non-monochromatic fields, we demonstrate that electromagnetic spin can undergo precessional dynamics. Important examples include spin precession in the rotating frame, for two orthogonally propagating circularly and linearly polarized plane waves, and the near-field of the precessing magnetic dipole.  The precessional behavior constitutes an optical analogue of the Landau–Lifshitz equation, a cornerstone of spin dynamics in condensed matter systems. For systems with magnetic sources, we derive the general photonic spin continuity equation \eqref{eq:spin_continuity_with_source} containing a source term in the form of the well-known spin torque, revealing a back-action exerted on the photonic spin from the magnetic sources and thus the complementary nature of spin in materials and electromagnetic fields. 
%\textcolor{red}{
Straightforward generalization to magnetic dipole chains suggests the full electromagnetic analogue of spin waves~-- the photonic spin waves~-- accompanying spin waves in matter, which sheds new light onto the electromagnetic part of magnon-polaritons.%} %, and could open ways to potential applications for photonic spintronics~\cite{}. 

We believe these results contribute a step towards unifying photonics and magnetization dynamics within a common theoretical framework %for spin precession, 
establishing a foundation for optical spintronics and enabling new pathways for spin-based photonic devices.

%The precession of electromagnetic spin arises from the breaking of fundamental symmetries. In particular, time-reversal symmetry is broken for rotating observers and in presence of magnetic fields, leading to electromagnetic spin precession. Boundary condition plays a vital role  %The current source breaks translational invariance making the spin density invariant under translation. 
%Additionally, bulk-inversion asymmetry and structural-inversion asymmetry give rise to effective magnetic fields that further drive the precessional motion of optical spin.

%When we apply the concept of precession of optical spin to magnetic material, we discover the existence of photonic spin wave.  

%This discovery 
%indings bridge photonics and magnonics, 
%establishing a unified paradigm for spin dynamics in electromagnetic systems and opening pathways for advanced optical spintronics and next-generation spin-based photonic devices.

\begin{acknowledgments}
We thank Konstantin Y. Bliokh and Ivan Fernandez-Corbaton for insightful discussions, critical feedback, and valuable suggestions that significantly improved this work.  The work has been supported by Russian Science Foundation (project 23-72-10059). A.A.B. acknowledges support from the National Natural Science Foundation of China (Project W2532010)
\end{acknowledgments}

%\section{References}

\bibliography{apssamp}% Produces the bibliography via BibTeX.

\onecolumngrid
\appendix
\section{Free time-harmonic waves do not precess}\label{Spin_Harmonic_waves}
We begin with the given definitions:
\begin{equation}
\tilde{\bm{E}} = -\frac{1}{\epsilon_0} \nabla\times\tilde{\bm{C}},\quad
\tilde{\bm{B}} = \nabla \times \tilde{\bm{A}}.
\end{equation}
For time-harmonic fields, the fields and potentials are represented as:
\begin{equation}
\bm{E}(\bm{r},t) = \operatorname{Re}[\tilde{\bm{E}}(\bm{r})e^{-i\omega t}],\quad
\bm{B}(\bm{r},t) = \operatorname{Re}[\tilde{\bm{B}}(\bm{r})e^{-i\omega t}],
\end{equation}
\begin{equation}
\bm{A}^\perp(\bm{r},t) = \operatorname{Re}[\tilde{\bm{A}}(\bm{r})e^{-i\omega t}],\quad
\bm{C}^\perp(\bm{r},t) = \operatorname{Re}[\tilde{\bm{C}}(\bm{r})e^{-i\omega t}].
\end{equation}

The spin density is:
\begin{equation}
\bm{s} = \frac{1}{2}[\epsilon_0\bm{E} \times \bm{A}^\perp + \bm{B} \times \bm{C}^\perp].
\end{equation}
Substituting the time-harmonic forms, we compute the cross products exactly. %Setting for simplicity, ${\bm{A}^\perp}^* = \bm{A}^* $ ( ${\bm{C}^\perp}^* = \bm{C}^* $) and
Using $\operatorname{Re}[z] = \frac{1}{2}(z + z^*)$, we expand:
\begin{equation}
\bm{E} \times \bm{A}^\perp  = \frac{1}{4}[\tilde{\bm{E}}e^{-i\omega t} + \tilde{\bm{E}}^*e^{i\omega t}] \times [\tilde{\bm{A}}e^{-i\omega t} + \tilde{\bm{A}}^*e^{i\omega t}].
\end{equation}
Expanding the cross product:
\begin{equation}
\bm{E} \times \bm{A}^\perp = \frac{1}{4}[\tilde{\bm{E}} \times \tilde{\bm{A}}\, e^{-2i\omega t} + \tilde{\bm{E}} \times \tilde{\bm{A}}^* + \tilde{\bm{E}}^* \times \tilde{\bm{A}} + \tilde{\bm{E}}^* \times \tilde{\bm{A}}^* e^{2i\omega t}].
\end{equation}
Similarly:
\begin{equation}
\bm{B} \times \bm{C}^\perp = \frac{1}{4}[\tilde{\bm{B}} \times \tilde{\bm{C}}\, e^{-2i\omega t} + \tilde{\bm{B}} \times \tilde{\bm{C}}^* + \tilde{\bm{B}}^* \times \tilde{\bm{C}} + \tilde{\bm{B}}^* \times \tilde{\bm{C}}^* e^{2i\omega t}].
\end{equation}

Using $\tilde{\bm{E}} = -\frac{1}{\epsilon_0} \nabla \times \tilde{\bm{C}}$ and $\tilde{\bm{B}} = \nabla \times \tilde{\bm{A}}$, we analyze the time-dependent terms. First for $\tilde{\bm{E}} \times \tilde{\bm{A}}$:
\begin{equation}
\tilde{\bm{E}} \times \tilde{\bm{A}} = -\frac{1}{\epsilon_0}[\nabla \times \tilde{\bm{C}}] \times \tilde{\bm{A}}.
\end{equation}
From Maxwell's equations in the frequency domain:
\begin{equation}
\nabla \times \tilde{\bm{E}} = i\omega\tilde{\bm{B}} = i\omega(\nabla \times \tilde{\bm{A}}).
\end{equation}
This implies:
\begin{equation}
\nabla \times (\tilde{\bm{E}} - i\omega\tilde{\bm{A}}) = 0 \implies \tilde{\bm{E}} = i\omega\tilde{\bm{A}} + \nabla\phi,
\end{equation}
where $\phi$ is a scalar potential. In the absence of charge source ($\nabla\phi = 0$), we have:
\begin{equation}
\tilde{\bm{E}} = i\omega\tilde{\bm{A}}.
\end{equation}
Substituting back into $\tilde{\bm{E}} \times \tilde{\bm{A}}$:
\begin{equation}
\tilde{\bm{E}} \times \tilde{\bm{A}} = (i\omega\tilde{\bm{A}}) \times \tilde{\bm{A}} = i\omega(\tilde{\bm{A}} \times \tilde{\bm{A}}) = 0,
\end{equation}
since the cross product of any vector with itself is identically zero.

For $\tilde{\bm{B}} \times \tilde{\bm{C}}$, from $\nabla \times \tilde{\bm{B}} =\mu_0 \epsilon_0 \partial_t \bm{E}= -\mu_0 \nabla \times \partial_t \bm{C} = i \mu_0 \omega\tilde{\bm{C}}$ (since $\bm{B} =-\mu_0\partial_t\bm{C}$ from Maxwell's equations):
\begin{equation}
\tilde{\bm{C}} = - \frac{i}{\mu_0 \omega} \tilde{\bm{B}}.
\end{equation}
Thus:
\begin{equation}
\tilde{\bm{B}} \times \tilde{\bm{C}} = - \frac{i}{\mu_0 \omega}  \tilde{\bm{B}} \times \tilde{\bm{B}} = 0.
\end{equation}

The spin density $\bm{S}$ becomes:
\begin{equation}
\bm{s} = \frac{1}{4}[\epsilon_0(\tilde{\bm{E}} \times \tilde{\bm{A}}^* + \tilde{\bm{E}}^* \times \tilde{\bm{A}}) + (\tilde{\bm{B}} \times \tilde{\bm{C}}^* + \tilde{\bm{B}}^* \times \tilde{\bm{C}})] + (\text{terms with } e^{\pm 2i\omega t}).
\end{equation}
From the above, we have:
\begin{equation}
\tilde{\bm{E}} \times \tilde{\bm{A}} = 0,\quad \tilde{\bm{E}}^* \times \tilde{\bm{A}}^* = 0,
\end{equation}
and:
\begin{equation}
\tilde{\bm{B}} \times \tilde{\bm{C}} = 0,\quad \tilde{\bm{B}}^* \times \tilde{\bm{C}}^* = 0.
\end{equation}

The time-dependent terms in $\bm{s}$ are:
\begin{equation}
\frac{1}{4}[\epsilon_0\tilde{\bm{E}} \times \tilde{\bm{A}}\, e^{-2i\omega t} + \epsilon_0\tilde{\bm{E}}^* \times \tilde{\bm{A}}^* e^{2i\omega t} + \tilde{\bm{B}} \times \tilde{\bm{C}}\, e^{-2i\omega t} + \tilde{\bm{B}}^* \times \tilde{\bm{C}}^* e^{2i\omega t}].
\end{equation}
Substituting $\tilde{\bm{E}} \times \tilde{\bm{A}} = 0$ and $\tilde{\bm{B}} \times \tilde{\bm{C}} = 0$:
\begin{equation}
\text{Time-dependent terms} = 0.
\end{equation}

After cancellation of the oscillatory terms, the spin density reduces to:
\begin{equation}
\bm{s} = \frac{1}{4}[\epsilon_0(\tilde{\bm{E}} \times \tilde{\bm{A}}^* + \tilde{\bm{E}}^* \times \tilde{\bm{A}}) + (\tilde{\bm{B}} \times \tilde{\bm{C}}^* + \tilde{\bm{B}}^* \times \tilde{\bm{C}})].
\end{equation}
Using $\tilde{\bm{E}} = i\omega\tilde{\bm{A}}$ and $\tilde{\bm{C}} = -\frac{i}{\mu_0 \omega} \tilde{\bm{B}}$, we can further simplify:
\begin{equation}
\bm{s} = \frac{1}{4}[-\epsilon_0(i\omega\tilde{\bm{A}}^*) \times \tilde{\bm{A}} + \epsilon_0(i\omega\tilde{\bm{A}}) \times \tilde{\bm{A}}^* + (\text{similar terms for } \bm{B} \text{ and } \bm{C})].
\end{equation}
\begin{equation}
    \bm{s} = \frac{1}{2}[ \epsilon_0(i\omega\tilde{\bm{A}}) \times \tilde{\bm{A}}^* +  (i\omega\tilde{\bm{B}}) \times \tilde{\bm{C}}^* ].
\end{equation}

The key result is that the time-dependent terms cancel exactly, leaving $\bm{S}$ time-independent.

\section{Spin density of instantaneous electromagnetic waves}\label{Spin_density_Instantaneous_Fields}
The spin angular momentum (SAM) of electromagnetic fields has been expressed in various forms in the literature. Stephen Barnett employed the duplex transformation, deriving the SAM for harmonic fields in vacuum as~\cite{Barnett2010,Barnett2011}
\begin{equation}
\bm{S} = \frac{1}{2} \int d^3 \bm{r} \left( \epsilon_0 \bm{E} \times \bm{A}^\perp + \bm{B} \times \bm{C} ^\perp\right),
\label{spin_harmonic}
\end{equation}
a result later referenced by Cameron~\cite{Cameron2012}. 

For more general cases, Andrea Aiello introduced the inverse curl operator to express the SAM solely in terms of observable fields, valid in free space without sources~\cite{Aiello2022}:
\begin{equation}
\bm{S} = \frac{\epsilon_0}{8 \pi c^2} \int d^3 \bm{r} \int d^3 \bm{r'} \frac{\bm{E}(\bm{r},t) \times \partial_t \bm{E}(\bm{r'},t) + c^2 \bm{B}(\bm{r},t) \times \partial_t \bm{B}(\bm{r'},t)}{|\bm{r} -\bm{r'}|}.
\label{formula1}
\end{equation}

An alternative formulation by Stewart, derived directly from Maxwell's equations and Helmholtz theory gives ~\cite{Stewart2005May},
\begin{equation}
\bm{S} = \frac{\epsilon_0}{4\pi} \int d^3 \bm{r} \int d^3 \bm{r'} \frac{\bm{B}(\bm{r},t) \times \partial_t \bm{B}(\bm{r'},t)}{|\bm{r} -\bm{r'}|}.
\label{formula2}
\end{equation}

While Eqs.~\eqref{formula1} and \eqref{formula2} are equivalent in free space~\cite{Aiello2022}, they differ in their domain of applicability: Eq.~\eqref{spin_harmonic} is restricted to harmonic fields, whereas Eq.~\eqref{formula1} holds for arbitrary instantaneous fields in free space. Equation~\eqref{formula2} has the broadest validity, applying to instantaneous fields even in the presence of current sources, but not for charge sources.

\section{Spin density analysis of transverse plane waves} \label{Appendix_superpositon_transverse_plane_waves} 
As a first step toward understanding the spin densities $\bm{s}_\mathrm{m}$ and $\bm{s}_\mathrm{e}$, we consider a superposition of transverse plane waves propagating along $\hat{\bm{z}}$:
\begin{equation}
    \begin{split}
        \bm{B} &= \sum_j \tilde{\bm{B}}_j e^{i(k_j z -\omega_j t)} \\
        \bm{E} &= \sum_j \tilde{\bm{E}}_j e^{i(k_j z -\omega_j t)}
    \end{split}
\end{equation}
where $j$ labels field components with amplitudes $\tilde{\bm{B}}_j$, $\tilde{\bm{E}}_j$, wavenumbers $k_j$, and frequencies $\omega_j$. The electric vector potential $\bm{C}^\perp$ in Coulomb gauge becomes:
\begin{equation}
    \begin{split}
        \bm{C}^\perp & = \frac{\epsilon_0}{4 \pi} \int d^3\bm{r} \frac{\partial \bm{B}(z',t)/\partial t}{|z -z'|} \\
        &= - i \epsilon_0 \sum_j \frac{\omega_j}{k^2_j}\tilde{\bm{B}}_j  e^{i(k_j z -\omega_j t)} \\
    \end{split}
\end{equation}
For transverse fields ($B_z = C_z = 0$), the second term in $\bm{s}_\mathrm{m}$ simplifies as:
\begin{equation}
    \begin{split}
        \left(\bm{B} \cdot \nabla \right) \left(\bm{r} \times \bm{C}^{\perp} \right) & = \left(B_x \, \frac{\partial}{\partial x} + B_y \,\frac{\partial}{\partial y} \right) \left[-z C_y \hat{\bm{x}} + z C_x \hat{\bm{y}} + \left(x C^\perp_y - y C^\perp _x\right) \hat{\bm{z}}\right]\\ & = \left(B_x \, \frac{\partial}{\partial x} + B_y \,\frac{\partial}{\partial y} \right) \left(x C^\perp_y - y C^\perp _x\right) \hat{\bm{z}} \\ \\
        & = B_x C_y \hat{\bm{z}} - B_y C_x \hat{\bm{z}}\\
        & = -i \epsilon_0 \left[ \sum_j \tilde{\bm{B}}_j e^{i(k_j z -\omega_j t)} \right]_x \left[ \sum_l \frac{\omega_l}{k_l^2}\tilde{\bm{B}}_l e^{i(k_l z -\omega_l t)} \right]_y \hat{\bm{z}} \\
        & +i \epsilon_0 \left[ \sum_j \tilde{\bm{B}}_j e^{i(k_j z -\omega_j t)} \right]_y \left[ \sum_l \frac{\omega_l}{k_l^2}\tilde{\bm{B}}_l e^{i(k_l z -\omega_l t)} \right]_x \hat{\bm{z}} \\ 
        & = \left( \bm{B} \times \bm{C}^{\perp} \right)_z \,\hat{\bm{z}}
    \end{split}
\end{equation}
 The spatial divergence term $(\bm{B} \cdot \nabla)(\bm{r} \times \bm{C}^\perp)$ exactly cancel the canonical spins' densities in plane-wave superpositions. Experimentally, however, the fields vanish at the boundaries and they do not contribute to the total spin \cite{Stewart2005May}. 

\section{Electromagnetic spin precession due to rotating observer}\label{PrecessionRotation}
%==================================================================
Let us see the effect of measuring the spin vector in a rotating reference frame. It is important to note that the dynamical laws are invariant under uniform translation, but they are not invariant under uniform rotation. Maxwell's equations in the rotating reference frame in Gaussian's units are 
\begin{eqnarray}
\nabla \cdot \bm{B}  &=& 0\\
\nabla \times \bm{E} &=& -\frac{\partial \bm{B}}{\partial t} \\
\nabla  \cdot \bm{E}  &=& \rho + \sigma \\
\nabla \times \bm{B} &=& \ \frac{\partial \bm{E}}{\partial t} + \bm{j} + \bm{i} 
\end{eqnarray}
where the extra terms \cite{Modesitt1970}
\begin{eqnarray}
\sigma  &=& \nabla \cdot (\bm{v} \times \bm{B}) = 2 \omega \cdot \bm{B} - \bm{v} \cdot (\nabla \times \bm{B}) \\
\bm{i} &=& \bm{v} \times (\nabla \times \bm{E}) + \nabla \times \left[ \bm{v} \times (\bm{E} - \bm{v} \times \bm{B}) \right]
\end{eqnarray}
The observer is moving at velocity $\bm{v} = \bm{\Omega} \times \bm{r}$ and $\bm{\Omega}$ is the angular velocity of rotation. Employing Stewart's formula (\ref{formula2}) in Gaussian's units, one can show, after lengthy algebra, that the spin vector in the rotating reference frame is
\begin{equation}
\bm{S}_{\rm{rot}} = \frac{B^2 V}{4 \pi \omega} \left( 1 - \frac{v_z}{c} \right) \hat{z} - \frac{B^2 V}{4 \pi \omega^2} \bm{\Omega}
\end{equation}
where $V$ is a normalization volume. Taking the time derivative in the rotating frame $\left( \frac{d \bm{S}}{dt}\right)_{\rm{rot}}$ is taken and assuming $\Omega_x =\Omega_y =0 $, one finds 
\begin{eqnarray}
    \left( \frac{d \bm{S}}{dt}\right)_{\rm{rot}} &=& 
    \frac{B^2 V}{4 \pi \omega} \left( 1 - \frac{\Omega_z}{\omega} \right) \left(\frac{d\hat{\bm{z}}}{dt}\right)_{\rm{rot}} \\
                                                 &=& \bm{S}_{\rm{rot}} \times \bm{\Omega} \\ 
                                                 &=& 
 \frac{\omega}{B} \bm{S}_{\rm{rot}} \times \frac{\bm{\Omega}}{\omega} B 
 \label{precessionrotating}
\end{eqnarray}
Equation (\ref{precessionrotating}) for optical spin exactly mirrors Landau-Lifshitz equation for precession of particles if we think of $\frac{\bm{\Omega}}{\omega} B$ as a fictitious magnetic field. 
%==================================================================
\section{Electromagnetic spin precession due to interference of a circularly-polarized wave with a constant magnetic field}\label{PrecessionBeff}
%==================================================================
\subsection{Orbital angular momentum of a circularly polarized wave superimposed over constant magnetic field}\label{orbital_angular_momentum_Bc_plus_B0}
The magnetic field of a circularly polarized wave superimposed over a static magnetic field directed along the $z$-axis is given by
\begin{equation}
    \bm{B}(\bm{r}, t) = B_c \cos(kz - \omega t) \hat{\bm{x}} 
                       + B_c \sin(k z -\omega t) \hat{\bm{y}} 
                       + B_0 \hat{\bm{z}},
\label{eq:Bc_plus_B0}
\end{equation}
where $B_c$ is the amplitude of the circularly polarized component, $B_0$ is the static field strength, $k$ is the wavenumber, and $\omega$ is the angular frequency. The electric vector potential is 
\begin{equation}
\begin{split}
 \bm{C}^{\perp} & = \frac{\epsilon_0}{4 \pi} \int \frac{d^3 r}{|\bm{r} - \bm{r}'|} \frac{\partial \bm{B}(\bm{r},t)}{\partial t} \\
 & =  \frac{\epsilon_0}{k^2} \frac{\partial \bm{B}_c(\bm{r},t)}{\partial t} 
\end{split}
\end{equation}
The orbital angular momentum (OAM) density associated with this field configuration can be computed as
\begin{equation}
    \begin{split}
        \int d^3r \, B_i (\bm{r} \times \nabla C^\perp_i) 
        &= \int d^3r \left[ B_x (\bm{r} \times \nabla C^\perp_x) 
                          + B_y (\bm{r} \times \nabla C^\perp_y) 
                          + B_z (\bm{r} \times \nabla C^\perp_z) \right] \\
        &= -\epsilon_0 \frac{B_c^2 \omega}{k} \int d^3r \, (-x \hat{\bm{y}} + y \hat{\bm{x}}) 
           \left[ \cos(kz - \omega t) + \sin(kz - \omega t) \right] \\
        &= 0,
    \end{split}
\end{equation}
where $C^\perp_i$ denotes the transverse component of the vector potential. 
As expected, the orbital angular momentum vanishes for this superposition of fields, 
consistent with the symmetry of the configuration.

\subsection{Spin of interference of circularly polarized wave with constant magnetic field}

The integral of the second term in the Belinfante's total angular moment vanishes when fields' profile end abruptly at the measuring device so that no surface effects contribute to the spin: 
\begin{equation}
    \int d^3r \, \bm{\nabla} \cdot \left[\bm{B} \otimes (\bm{r} \times \bm{C}^\perp) \right] = \oint_\Gamma d 
    \bm{\Gamma} \cdot \bm{B} \, (\bm{r} \times \bm{C}^\perp) =0,
\end{equation}
Hence, the Belinfante's total angular momentum reduces to the canonical magnetically-biased SAM of non-monochromatic fields 
\begin{equation}
\bm{S} = \frac{1}{(4 \pi c)^2} \int d^3 \bm{x} \int d^3 \bm{x}' \frac{\bm{B}(\bm{x},t)}{|{\bm{x} - \bm{x}'}|} \times \frac{\partial \bm{B}(\bm{x}',t)}{\partial t} 
\label{stewart}
\end{equation}
reducing to the  the general spin expression derived by Steward based on Helmeholtz's decomposition theorem \cite{Stewart2005May}. The magnetic field is the sum of two fields: the magnetic field of circularly polarized wave $\bm{B}_c(\bm{r},t) $  and and a constant magnetic field $\bm{B}_0$ Eq. (\ref{eq:Bc_plus_B0}).  
Using the relations \cite{Stewart2005May}
\begin{equation}
\int d^3r' \, \frac{\cos \mathbf{k} \cdot \mathbf{r}'}{|\mathbf{r} - \mathbf{r}'|} = \frac{4\pi}{k^2} \cos \mathbf{k} \cdot \mathbf{r}
\label{eq:idenity_1}
\end{equation}
and
\begin{equation}
\int d^3r' \, \frac{\sin \mathbf{k} \cdot \mathbf{r}'}{|\mathbf{r} - \mathbf{r}'|} = \frac{4\pi}{k^2} \sin \mathbf{k} \cdot \mathbf{r}
\label{eq:idenity_2}
\end{equation}
When we substitute Eq.~(\ref{eq:Bc_plus_B0}) in Eq.~(\ref{stewart}), it follows that the spin vector is 
\begin{equation}
\bm{S} = a \cos(\omega t) \hat{\bm{x}} + a  \sin(\omega t) \hat{\bm{y}} + b \hat{\bm{z}}
\label{spinvolume}
\end{equation}

where $a = \epsilon_0 \frac{8 B_c B_0  \sin(kl)}{k^3} L^2$ and $b = \epsilon_0 \left(\frac{B_c}{k}\right)^2 \omega L^3 \hat{\bm{z}}$. It follows directly that the electromagnetic spin Eq.(\ref{spinvolume}) obeys the following differential equation
\begin{equation}
\frac{d \bm{S}}{d t} =  \omega \, \bm{S} \times \hat{z}
\label{lightprecession}
\end{equation}
Equation (\ref{lightprecession}) for electromagnetic spin mirrors Landau-Lifshitz equation for the magnetization dynamics in magnetic materials. One possible implication of eq.(\ref{lightprecession}) is that an externally applied magnetic field exerts torque on optical spin setting the latter into precessional motion.

\subsection{Electrically-biased SAM: circularly polarized wave and constant magnetic field}\label{Appendix:electrically-biased-formulation-constant-magnetic-field}
The electric field corresponding to the magnetic field configuration in Eq.(\ref{eq:Bc_plus_B0}) is 
\begin{equation}
    \bm{E} = E_0 \sin( \omega t - k z) \hat{\bm{x}} +  E_0 \cos(\omega t -k z) \hat{\bm{y}}
\end{equation}
The magnetic vector potential is 
\begin{equation}
    \begin{split}
        \bm{A}^\perp & = \bm{A}^\perp_\text{c} + \bm{A}^\perp_0 \\
        & = - A_\text{c} \left[\cos(kz - \omega t) \hat{\bm{x}} + \sin(kz - \omega t) \hat{\bm{y}} \right] + \frac{1}{2} B_0 \left(y \hat{\bm{x}} -x \hat{\bm{y}}  \right)
    \end{split}
\end{equation}
where $\bm{\nabla} \cdot \bm{A} = 0$ in the Coulomb gauge. The canonical spin is 
\begin{equation}
\begin{split}
\bm{S}_{\rm{e}}^{\rm{can}} & = \epsilon_0 \int \bm{E} \times \bm{A}\, d^3r \\ 
& = \left(\frac{B_0}{k} \right)^2 \omega L^3 \hat{\bm{z}}
\end{split}
\end{equation}

The canonical orbital angular momentum is 
\begin{equation}
\begin{split}
    \bm{L}_{\rm{e}}^{\rm{can}} & =    \int d^3r E_i \left(\bm{r} \times \bm{\nabla} \right) A_i \\ \\
    & = \int d^3r\begin{bmatrix} 
   y A_c E_0 k - \frac{1}{2} z B_0 E_0  \sin(\omega t - k z)\\
- x A_c E_0 k + \frac{1}{2} z B_0 E_0  \cos( \omega t - k z) \\
\frac{1}{2} B_0 E_0 \left(x  \sin(\omega t - kz) -y \cos( \omega t- k z) \right)
\end{bmatrix} \\ 
\\
& = a_1 \cos(\omega t) \hat{\bm{x}} + a_1 \sin(\omega t)  \hat{\bm{y}} + b \hat{\bm{z}}
\end{split}
\end{equation}
where $a_1 = 8 B_0 E_0 L^2 \big(-kL \cos(kL) + \sin(kL)\big)/k^2
$ in the Coulomb gauge. 
In contrast to the magnetic-bias formulation which is gauge independent, the electric-biased formulation is gauge dependent for this electromagnetic field configuration.

\section{Electromagnetic spin precession due to mixing of a circularly polarized wave and a linearly-polarized wave}\label{App_Precessiing_Plane_Waves}

%\subsection{The magnetically biased formulation is gauge-independent}
Consider the superposition between a circularly polarized wave and a plane wave with different frequencies:

\begin{equation}
\bm{B} = \bm{B}_\text{c} + \bm{B}_\text{l}
\end{equation}

\begin{equation}
\bm{B} = \left[B_\text{c} \cos( k_1 z - \omega_1 t) \hat{\bm{x}} + B_\text{c} \sin(k_1 z - \omega_1 t ) \hat{\bm{y}}\right] + B_\text{l} \cos( k_2 x - \omega_2 t) \hat{\bm{z}}
\end{equation}
The electric vector potential can be calculated as 
\begin{equation}
\begin{split}
 \bm{C}^{\perp} & = \frac{\epsilon_0}{4 \pi} \int \frac{d^3 r}{|\bm{r} - \bm{r}'|} \frac{\partial \bm{B}(\bm{r},t)}{\partial t} \\
 & = \frac{\epsilon_0}{4 \pi} \left[\frac{4 \pi}{k_1^2} \frac{\partial \bm{B}_\text{c}(\bm{r},t)}{\partial t} + \frac{4 \pi}{k_2^2} \frac{\partial \bm{B}_\text{l}(\bm{r},t)}{\partial t} \right]
\end{split}
\end{equation}
where we made use of identities Eq.(\ref{eq:idenity_1}-\ref{eq:idenity_2}). Note that there is no ambiguity concerning the electric vector potential in this formulation. 

Assuming that waves have fields profiles that die abruptly at the measuring device boundaries, the total spin becomes
\begin{equation}
\begin{split}
\bm{S} &=  \frac{\epsilon_0}{4\pi}  \int d^3 \bm{r} \left[ 
\frac{4\pi}{k_1^2} \bm{B}_\text{c}(\bm{r}, t) \times \frac{\partial \bm{B}_\text{c}(\bm{r}, t)}{\partial t} + 
\frac{4\pi}{k_2^2} \bm{B}_\text{c}(\bm{r}, t) \times \frac{\partial \bm{B}_\text{l}(\bm{r}, t)}{\partial t} \right. \\
&\quad \left. + \frac{4\pi}{k_1^2} \bm{B}_\text{l}(\bm{r}, t) \times \frac{\partial \bm{B}_\text{c}(\bm{r}, t)}{\partial t} + 
\frac{4\pi}{k_2^2} \bm{B}_\text{l}(\bm{r}, t) \times \frac{\partial \bm{B}_\text{l}(\bm{r}, t)}{\partial t} \right]
\end{split}
\end{equation}
The last term is vanishes due to basic properties of vector products and 
\[
\frac{\epsilon_0}{4\pi} \frac{4\pi}{k_1^2} \int d^3 \bm{r} \, 
\bm{B}_\text{c}(\bm{r}, t) \times \frac{\partial \bm{B}_\text{c}(\bm{r}, t)}{\partial t}
= \frac{B_\text{c}^2}{k_1^2} L^3 \bm{\hat{z}}
\]
is the spin of a circularly polarized wave normalized in volume $V = \left(2 L \right)^3$.

The spin density of the two wave (neglecting the boundary term) is the sum of $\bm{s}_1$ and $\bm{s}_2$ 
\begin{equation}
\bm{s}_1 = \frac{\epsilon_0}{2} B_cB_l
\begin{bmatrix}
-\dfrac{\omega_1}{k_1^2} \sin\!\left[ \left(\omega_1 + \omega_2 \right)t - \left(k_1 z + k_2 x \right) \right] 
+ \dfrac{\omega_2}{k_2^2} \cos\!\left[ \left(\omega_1 + \omega_2 \right)t - \left(k_1 z + k_2 x \right) \right] \\[6pt]
\dfrac{\omega_2}{k_2^2} \sin\!\left[ \left(\omega_1 + \omega_2 \right)t - \left(k_1 z + k_2 x \right) \right] 
- \dfrac{\omega_2}{k_2^2} \sin\!\left[ \left(\omega_1 + \omega_2 \right)t - \left(k_1 z + k_2 x \right) \right] \\[6pt]
0
\end{bmatrix}
\end{equation} 

and 

\begin{equation}
\bm{s}_2 = \frac{\epsilon_0}{2} B_cB_l
\begin{bmatrix}
\dfrac{\omega_1}{k_1^2} \sin\!\left[ \left(\omega_2 - \omega_1 \right)t - \left(k_2 x - k_1 z \right) \right] 
- \dfrac{\omega_2}{k_2^2} \cos\!\left[ \left(\omega_2 - \omega_1 \right)t - \left(k_2 x - k_1 z \right) \right] \\[6pt]
-\dfrac{\omega_2}{k_2^2} \sin\!\left[ \left(\omega_2 - \omega_1 \right)t - \left(k_1 z - k_2 x \right) \right] 
+ \dfrac{\omega_2}{k_2^2} \sin\!\left[ \left(\omega_2 - \omega_1 \right)t - \left(k_2 x - k_1 z \right) \right] \\[6pt]
b
\end{bmatrix}
\end{equation}

The total spin can be decomposed to 
\begin{equation}
    \begin{aligned}
\bm{S}_1 &= \frac{a}{2} \left(k_2^2 \omega_1 - k_1^2 \omega_2\right) \cos\left((\omega_1 + \omega_2)t\right) \bm{\hat{x}} 
- \frac{a}{2} \left(k_2^2 \omega_2 - k_1^2 \omega_1\right) \sin\left((\omega_1 + \omega_2)t\right) \bm{\hat{y}} 
 \\ %+ \frac{b}{2} \bm{\hat{z}} \\
\bm{S}_2 &= \frac{a}{2} \left(k_2^2 \omega_1 + k_1^2 \omega_2\right) \cos\left((\omega_1 - \omega_2)t\right) \bm{\hat{x}} 
- \frac{a}{2} \left(k_1^2 \omega_2 + k_2^2 \omega_1\right) \sin\left((\omega_1 - \omega_2)t\right) \bm{\hat{y}} 
+ %\frac{b}{2} 
b \bm{\hat{z}}
\end{aligned}
\end{equation}
The constants are given by 
\[
a = \epsilon_0 \frac{8 B_\text{c} B_\text{l} \sin(k_1 L) \sin(k_2 L) L}{k^3_1   k^3_2}, \quad 
b = \epsilon_0 \left(\frac{B_\text{c}}{k_1}\right)^2 \omega_1 \left(2L\right)^3
\]
It is straightforward to show that the spins obey the following precession differential equations
\begin{equation}
    \begin{aligned}
\frac{d \bm{S}_1}{d t} &= \left( \omega_1 + \omega_2 \right) \bm{S}_1 \times \bm{\hat{z}}, \\
\frac{d \bm{S}_2}{d t} &= \left( \omega_1 - \omega_2 \right) \bm{S}_2 \times \bm{\hat{z}},
\end{aligned}
\end{equation} 
In the quasimonochromatic limit $\delta \omega = \omega_1 - \omega_2$, where $\delta \omega/\left(\omega_1 + \omega_2 \right) \ll 1$, one finds 
\begin{equation}
    \frac{d \bm{S}}{d t} = \delta \omega \, \bm{S} \times \hat{\bm{z}} + 2 \omega_2 \bm{S}_1 \times \bm{\hat{z}}, 
\end{equation}
That is the total spin $\bm{S}$ slowly precess at frequency $\delta \omega$. Meanwhile, small nutations occurs at a fast frequency $\omega_1 + \omega_2$.

%\subsection{Electrically-biased Balifente angular momentum: Mixing of orthogonal circularly polarized and linearly polarized waves}

%\section{Electromagnetic spin wave} 

\section{Continuity equations for instantaneous spin densities in the presence of current sources} \label{Appendix_continuity_equation}
The local spin continuity equation for free space harmonic fields is \cite{Bliokh2014,Alexeyev1999}:

\begin{equation}
    \partial_t S_i + \partial_j \Sigma_{ij} = 0,
\end{equation}

\begin{equation}
    S_i = \frac{1}{2} \left( \epsilon_0 \mathbf{E} \times \mathbf{A} +  \mathbf{B} \times \mathbf{C} \right)_i,
\end{equation}

\begin{equation}
    \Sigma_{ij} = \frac{1}{2} \left[ \delta_{ij} \left( \mathbf{B} \cdot \mathbf{A} - \mathbf{E} \cdot \mathbf{C} \right) - B_i A_j - B_j A_i + E_i C_j + E_j C_i \right], 
    \label{spin_current_nonharmonic}
\end{equation}

%This spin continuity equation does not account for the presence of sources. 
%\subsection{Spin continuity equation with current sources for instantaneous spin density} \label{continuity_equation_with_sources}
%\subsubsection*{A. Basics from Stewart's and Nienhuis's articles}
Helmholtz's decomposition theory states that any vector field that vanishes at infinity can be resolved into the sum of an irrotational (curl-free) vector field and a solenoidal (divergence-free) vector field \cite{Stewart2005May}
\begin{equation}
\bm{E}(\bm{x}, t) = -\nabla f(\bm{r}, t) - \nabla \times \bm{C}^{\perp}(\bm{r}, t) 
\end{equation} 
The electric field is decomposed into longitudinal part $\bm{E}^\parallel = -\nabla f(\bm{x}, t)  $ and transverse part $\bm{E}^\perp =  - \nabla \times \bm{C}^\perp(\bm{x}, t)$. The field  $\bm{E}^\parallel$ is then identical to the Coulomb field corresponding to the instantaneous charge distribution, while the transverse part $\bm{E}^\perp$ may be viewed as "the radiative electric field" or transverse electric field \cite{Nienhuis2016} %(\textcolor{green}{this is not entirely correct because in the absence of electric charge the electric field is purely transverse even in the near-field region})
. Employing Maxwell's equations \cite{Stewart2005May}
\begin{align}
 \bm{E}^\parallel = - \nabla f(\bm{r}, t) &=  - \nabla \int d^3\bm{r}' \frac{\rho(\bm{r}', t)}{4 \pi \epsilon_0 |\bm{r} - \bm{r}'|} \\
\bm{E}^\perp = - \frac{1}{\epsilon_0}\nabla \times \bm{C}^{\perp}(\bm{r}, t) &=- \nabla \times \int d^3 \bm{r}' \frac{\nabla'\times\bm{E}(\bm{r}', t)}{4 \pi \epsilon_0 |\bm{r} - \bm{r}'|} = \nabla \times  \int d^3 \bm{r}' \frac{\frac{\partial}{\partial t}\bm{B}(\bm{r}', t)}{4 \pi \epsilon_0 |\bm{r} - \bm{r}'|} 
\end{align} 
The magnetic vector potential is a nonlocal function of the magnetic field  \cite{Bialynicki2014,Stewart2003}
\begin{equation}
\bm{A}^\perp(\bm{r}, t) =  \nabla \times \int d^3\bm{r}' \frac{\bm{B}(\bm{r}', t)}{4 \pi |\bm{r} -\bm{r}'|}
\end{equation}
%\subsubsection*{B. Another Equation for spin density in presence of current source}

It has been shown that \cite{Aiello2022,Belinfante1962} 
\begin{equation}
\begin{split}
\nabla \times \left[ (\nabla \times)^{-1} \mathbf{G}^{\perp}(\mathbf{r}, t) \right] &= \nabla \times 
\left( \nabla \times \left[ \int \frac{d^3\mathbf{r}'}{4\pi} \frac{\mathbf{G}^{\perp}(\mathbf{r}', t)}{|\mathbf{r} - \mathbf{r}'|} \right] \right) \\
&= \nabla \left( \nabla \cdot \left[ \int \frac{d^3\mathbf{r}'}{4\pi} \frac{\mathbf{G}^{\perp}(\mathbf{r}', t)}{|\mathbf{r} - \mathbf{r}'|} \right] \right) 
- \nabla^2 \left[ \int \frac{d^3\mathbf{r}'}{4\pi} \frac{\mathbf{G}^{\perp}(\mathbf{r}', t)}{|\mathbf{r} - \mathbf{r}'|} \right] \\ 
&= -\nabla^2 \left[ \int \frac{d^3\mathbf{r}'}{4\pi} \frac{\mathbf{G}^{\perp}(\mathbf{r}', t)}{|\mathbf{r} - \mathbf{r}'|} \right] \\
&= \int d^3\mathbf{r}'\, \mathbf{G}^{\perp}(\mathbf{r}', t) \delta(\mathbf{r} - \mathbf{r}') \\
&= \mathbf{G}^{\perp}(\mathbf{r}, t)     
\end{split}
\label{InverseCurl}
\end{equation}

%\subsubsection*{C. Continuity equation and spin current}
Instantaneous spin density is 
\begin{equation}
\begin{split}
    \bm{s}_m & = \bm{B} \times \bm{C}^{\perp} - \left( \bm{B} \cdot \nabla \right) \left(\bm{r} \times \bm{C}^{\perp} \right) \\ 
    & = \bm{B} \times \bm{C}^{\perp} - \nabla \cdot \bm{ \mathsf{T}}
\end{split}
\end{equation}
where the tensor is given by $\bm{\mathsf{T}} = \bm{B} \otimes \left( \bm{r} \times \bm{C}^{\perp}\right)$. 

Taking the time derivative of the previous equation
\begin{equation}
    \frac{\partial \bm{s}}{\partial t} =  \frac{\partial\bm{B}(\bm{r},t)}{\partial t} \times \bm{C}(\bm{r},t) + \bm{B}(\bm{r},t) \times \frac{\partial \bm{C}(\bm{r},t)}{\partial t} -\left( \frac{\partial\bm{B}(\bm{r},t)}{\partial t} \cdot \nabla \right) \left(\bm{r} \times \bm{C}_{\perp}(\bm{r},t) \right) -\left( \bm{B}(\bm{r},t)\cdot \nabla \right) \left(\bm{r} \times \frac{\partial\bm{C}_\perp(\bm{r},t)}{\partial t} \right)
    \label{eq:continuity_equation_1}
\end{equation}
From Maxwell's equation: 
\begin{equation}
    \frac{\partial\bm{B}(\bm{r},t)}{\partial t} = -\nabla \times \bm{E}(\bm{r},t)
    \label{eq:continuity_equation_2}
\end{equation} 
\begin{equation}
\begin{split}
     \frac{\partial \bm{C}_\perp(\bm{r},t)}{\partial t} & =  -\frac{1}{\mu_0} \bm{B}(\bm{r},t) +  \int \frac{d^3\bm{r}'}{4 \pi |\bm{r} - \bm{r}'|} \nabla' \times \bm{J}(\bm{r}_0,t) \\
     & =- \frac{1}{\mu_0} \bm{B}(\bm{r},t) +  \nabla \times \int \frac{d^3\bm{r}'}{4 \pi |\bm{r} - \bm{r}'|} \times \bm{J}(\bm{r}_0,t)  \\ 
     & = - \frac{1}{ \mu_0} \bm{B}(\bm{r},t) +  \left(\nabla \times\right)^{-1} \bm{j}(\bm{r}_0,t) \\ 
     & =  - \frac{1}{ \mu_0} \bm{B}(\bm{r},t) +  \bm{M} \\ 
     \label{eq:continuity_equation_3}
\end{split}
\end{equation}
where $\bm{M}$ is the magnetization and $\bm{j}$ is the current source %and $\bm{H} = - \frac{1}{ \mu_0} \bm{B} +  \bm{M}$ is the magnetic field strength. \textcolor{magenta}{Check with Max the proper definition of the current source position and coordinates of magnetization.} %\textcolor{red}{here perhaps a misprint in the last line}

Substituting Eqs.(\ref{eq:continuity_equation_2}-\ref{eq:continuity_equation_3}) in Eq.(\ref{eq:continuity_equation_1}) and using some general vector and tensor relations, it is straightforward to arrive at the following equation for the spin density of instantaneous fields in presence of current sources \\ 

{\centering
\fcolorbox{blue}{white}{%
$\displaystyle
    \frac{\partial \bm{s}_{\mathrm{m}}}{\partial t}  +
    \nabla \cdot \left(\bm{\mathsf{\Sigma}} + \frac{\partial \bm{ \mathsf{T}}}{\partial t} \right)  
    = - \bm{M} \times \bm{B} 
    \label{Eq_derivative_sm}
$%
}\par} \
%\noindent where $\bm{r}_0$ marks the center of the current distribution. \\
\\
The current source in the continuity equation is general, for magnetization current $\bm{j} = \nabla \times \bm{M}$, it is straightforward to see that the right hand side of the continuity equation corresponds to the magnetization torque density $ - \bm{M} \times \bm{B}$. This term in the spin current $\bm{\mathsf{\Phi}} = \frac{\partial \bm{ \mathsf{T}}}{\partial t}$ represents an additional contributions for non-harmonic fields.

The instantaneous spin current $\bm{\mathsf{\Pi}}$ is:
\begin{equation}
\begin{split}
    \bm{\mathsf{\Pi}}  & = \bm{\mathsf{\Sigma}} +  \bm{\mathsf{\Phi}} \\
    & = \bm{E} \otimes \bm{C}^{\perp} + \bm{C}^{\perp} \otimes \bm{E} - \left( \bm{E} \cdot \bm{C}^{\perp} \right) \bm{\mathsf{I}} + \frac{\partial \bm{ \mathsf{T}}}{\partial t}  \\ 
    & = \bm{E} \otimes \bm{C}^{\perp} + \bm{C}^{\perp} \otimes \bm{E} - \left( \bm{E} \cdot \bm{C}^{\perp} \right) \bm{\mathsf{I}} + \frac{\partial \bm{B}}{\partial t} \otimes \left( \bm{r} \times \bm{C}^{\perp}\right) +\bm{B} \otimes \left( \bm{r} \times \frac{\partial \bm{C}^{\perp}}{\partial t}\right) \\  
 &= \bm{E} \otimes \bm{C}^{\perp} + \bm{C}^{\perp} \otimes \bm{E} - \left( \bm{E} \cdot \bm{C}^{\perp} \right) \bm{\mathsf{I}} +  \frac{\partial \bm{B}}{\partial t} \otimes \left( \bm{r} \times \bm{C}^{\perp}\right)- \left[\frac{1}{\mu_0}  \bm{B} \left( \bm{r} \times \bm{B}\right) -  \bm{B} \otimes \left( \bm{r} \times \bm{M}\right) \right] \\
 %& =  \bm{E} \otimes \bm{C}_{\perp} + \bm{C}_{\perp} \otimes \bm{E} - \left( \bm{E} \cdot \bm{C}_{\perp} \right) \bm{\mathsf{I}} +  \frac{\partial \bm{B}}{\partial t} \otimes \left( \bm{r} \times \bm{C}^{\perp}\right) - \bm{B} \otimes \left(\bm{r} \times \bm{H} \right) 
\end{split}
\end{equation}

The total spin current $ \bm{\bm{\mathsf{\Pi}}}$ has the following physical features: 
\begin{itemize}
    \item Translation invariance in case of current source. In case of a magnetic dipole, the current is 
\begin{equation}
\bm{j} = \nabla \times \bm{m} \, \delta\left(\bm{r}\right)
\end{equation}
\begin{equation}
    \bm{\mathsf{\Pi}}  = \bm{E} \otimes \bm{C}^{\perp} + \bm{C}^{\perp} \otimes \bm{E} - \left( \bm{E} \cdot \bm{C}^{\perp} \right) \bm{\mathsf{I}} +  \frac{\partial \bm{B}}{\partial t} \otimes \left( \bm{r} \times \bm{C}^{\perp}\right)- \left[\frac{1}{\mu_0}  \bm{B} \left( \bm{r} \times \bm{B}\right) - \left( \frac{\mu_0}{4 \pi}\right) \frac{8 \pi}{3}  \bm{B} \otimes \left( \bm{r} \times \bm{m} \, \delta\left( \bm{r} \right)  \right) \right] 
\end{equation}
Translation of the observation point $\bm{r}$ by a vector $\bm{r_0}$ is 
\item In case of harmonic fields, 
\[
\langle \bm{\mathsf{\Phi}} \rangle = 0
\]
as the tensor $\bm{ \mathsf{T}}$ is periodic. The spin current reduces to the time harmonic expression $\langle \bm{\mathsf{\Pi}} \rangle = \langle\bm{\mathsf{\Sigma}} \rangle$
\end{itemize}

%The last line in Eq. (13) shows that as long as $\bm{B}$ and 
%\textcolor{red}{need to think what should be substituted as $\bm{B}$ here~-- total time-dependent field or only its time-constant components + external constant magnetic field, or just the external constant magnetic field? because I think we should avoid self-interaction terms in our approach, I think the latter. Need to check in derivation} 
%$\bm{J} = 4 \pi \nabla \times (\bm{m} \cdot \delta (\bf{r}-\bf{r_0}))$
%$ -  \delta (\bf{r}-\bf{r_0}) \cdot \bm{m}\times \bm{B} $

\section{Electromagnetic spin precessional motion from a single magnetic dipole} \label{spin_precession_dipole}
Assume a precessing magnetic dipole whose magnetic moment is given by
\begin{equation}
    \bm{m} = m_0 \cos\left( \omega t\right) \hat{x} + m_0 \sin\left( \omega t\right) \hat{y} + b_0 \hat{z}
\end{equation}
The magnetic field of an arbitrary time varying magnetic moment of a point dipole is given by 
\begin{equation}
    \bm{B}(\bm{r},t)  = - \frac{\mu_0}{4 \pi}  \left( \frac{\bm{m}  - 3 \bm{n} \left(\bm{n} \cdot \bm{m}  \right) }{R^3} +   \frac{\dot{\bm{m}}  - 3 \bm{n} \left(\bm{n} \cdot \dot{\bm{m}}  \right) }{c R^2} + \frac{\ddot{\bm{m}}  - \bm{n} \left(\bm{n} \cdot \ddot{\bm{m}}  \right) }{c^2 R} - \frac{8 \pi}{3} \bm{m} \, \delta(\bm{r} - \bm{Z}_0) \right)
\end{equation}
where $\bm{R} = \bm{r} - \bm{Z_0}$ with $\bm{R}$ and $\bm{Z_0}$ being the field and dipole positions, respectively. The unit vector $\bm{n}$ is in the direction of the vector $\bm{R}$. The magnetic moment should be evaluated at the retarded time $t_r = t -R/c$. However, retardation effects are negligible in the near-field zone. The first term is the magnetic field in the near field region, the second term is in the mid-field zone, and the third is in the far zone. The fourth term is important and it represents the magnetic field inside the magnetized source. This term is required to give correct field values at the origin. The last term is the magnetic field inside the dipole \cite{Heras1998}. We can set the dipole at the origin of coordinate system in this case $\bm{Z_0} = 0$

It follows that spin of the magnetic dipole under arbitrary temporal variation can be expressed as 
\begin{equation}
    \begin{split}
        \bm{S} &= \frac{\epsilon_0}{4 \pi} \int d^3 \bm{r}  \int d^3 \bm{r'} 
        \Biggl[ 
        \frac{\bm{B}_{\rm{NF}}(\bm{r},t) \times \frac{\partial \bm{B}_{\rm{NF}}(\bm{r'},t) }{\partial t}  }{|\bm{r} -\bm{r'}|} 
        +  \frac{\bm{B}_{\rm{NF}}(\bm{r},t) \times \frac{\partial \bm{B}_{\rm{MF}}(\bm{r'},t) }{\partial t}  }{|\bm{r} -\bm{r'}|} + \frac{\bm{B}_{\rm{NF}}(\bm{r},t) \times \frac{\partial \bm{B}_{\rm{FF}}(\bm{r'},t) }{\partial t}  }{|\bm{r} -\bm{r'}|} 
        \\ 
        &\quad +  \frac{\bm{B}_{\rm{MF}}(\bm{r},t) \times \frac{\partial \bm{B}_{\rm{NF}}(\bm{r'},t) }{\partial t}  }{|\bm{r} -\bm{r'}|} + \frac{\bm{B}_{\rm{MF}}(\bm{r},t) \times \frac{\partial \bm{B}_{\rm{MF}}(\bm{r'},t) }{\partial t}  }{|\bm{r} -\bm{r'}|} 
        +  \frac{\bm{B}_{\rm{MF}}(\bm{r},t) \times \frac{\partial \bm{B}_{\rm{FF}}(\bm{r'},t) }{\partial t}  }{|\bm{r} -\bm{r'}|} \\ 
        &\quad + \frac{\bm{B}_{\rm{FF}}(\bm{r},t) \times \frac{\partial \bm{B}_{\rm{NF}}(\bm{r'},t) }{\partial t}  }{|\bm{r} -\bm{r'}|} 
        +  \frac{\bm{B}_{\rm{FF}}(\bm{r},t) \times \frac{\partial \bm{B}_{\rm{MF}}(\bm{r'},t) }{\partial t}  }{|\bm{r} -\bm{r'}|} 
        +  \frac{\bm{B}_{\rm{FF}}(\bm{r},t) \times \frac{\partial \bm{B}_{\rm{FF}}(\bm{r'},t) }{\partial t}  }{|\bm{r} -\bm{r'}|} 
        \Biggr] 
    \end{split}
\end{equation}

The value of the magnetic field in the mid-field and far-field regions are proportional to $(\frac{\omega}{c})$ and $(\frac{\omega}{c})^2$, respectively. In GHz range, $\frac{\omega}{c} \approx 10^{-4} \, cm$, hence, the near-field electromagnetic fields dominate.  
\begin{equation}
\begin{split}
     \bm{B}(\bm{r},t)  & = - \frac{\mu_0}{4 \pi}  \frac{\bm{m}  - 3 \hat{\bm{r}} \left(\hat{\bm{r}} \cdot \bm{m}  \right) }{r^3} + \left(\frac{\mu_0}{4 \pi} \right) \frac{8 \pi}{3} \bm{m} \, \delta(\bm{r})   \label{A3}
\end{split} 
\end{equation}
\\

\section{Magnetically-biased angular momentum density of a single precessing dipole}\label{Appendix_Magnetically-biased_spin_density}

\subsection{Magnetically-biased orbital angular momentum} \label{Appendix:Orbital_angular_momentum}

We first evaluate the magnetically-biased orbital angular momentum for a general electromagnetic field:
\begin{equation}
\bm{L}_{\rm{m}} = \frac{\epsilon_0}{4\pi} \int d^3r \int d^3r' \, \left[ \bm{B}(\bm{r}, t) \cdot  \frac{\partial \bm{B}(\bm{r}', t)}{\partial t} \right]  \frac{\bm{r} \times \bm{r}'}{|\bm{r} - \bm{r}'|^3}
\label{eq:Lm_general}
\end{equation}

For a precessing magnetic dipole in the near-field regime, Eq.~\eqref{eq:Lm_general} reduces to:
\begin{equation}
\begin{split}
\bm{L}_{\rm{m}}
    & = \frac{\epsilon_0}{4 \pi} {\left(\frac{\mu_0}{4 \pi} \right)}^2 \int d^3 r\int d^3 r' \frac{\bm{r} \times \bm{r}'}{|\bm{r} - \bm{r'}|} \big\{ \\
& \frac{1}{r^3 r'^3}\left[ \bm{m} \times \dot{\bm{m}} + 3 (\hat{\bm{r}}' \times \bm{m}) (\hat{\bm{r}}' \cdot \dot{\bm{m}}) - 3 (\hat{\bm{r}} \times \dot{\bm{m}}) (\hat{\bm{r}} \cdot \bm{m})  + 9 (\hat{\bm{r}} \times \bm{\bm{r}}') (\hat{\bm{r}} \cdot \bm{m})(\hat{\bm{r}}' \cdot \dot{\bm{m}}) \right]  \\ 
    &  - \frac{8 \pi}{3}\frac{1}{r'^3} \left[ \left(\bm{m} \cdot \dot{\bm{m}} \right) \delta\left(\bm{r}\right) -3 (\hat{\bm{r}}' \cdot \bm{m})^2 \delta\left(\bm{r}'\right)   \right] - \frac{8 \pi}{3}\frac{1}{r^3} \left[ \left(\bm{m} \cdot \dot{\bm{m}} \right) \delta\left(\bm{r}'\right) -3 (\hat{\bm{r}} \cdot \dot{\bm{m}}) (\hat{\bm{r}} \cdot \bm{m}) \delta\left(\bm{r}'\right)   \right]\\
    & + \left(\frac{8 \pi}{3}\right)^2    \left(\bm{m} \cdot \dot{\bm{m}} \right) \,\delta\left(\bm{r}\right) \delta\left(\bm{r}'\right) \big\} 
\end{split}
\end{equation}
Second line of the equation 
\begin{itemize}
    \item The first term vanishes due to azimuth symmetry 
    \item The third term vanishes due to azimuth symmetry 
    \item The components of the second and fourth integrals cancel each others.
\end{itemize}
All Dirac-$\delta$ terms integrate to zero. 

Thus, $\bm{L}_{\mathrm{m}} = 0$, demonstrating that a precessing magnetic dipole generates \emph{pure spin} angular momentum in the gauge-independent magnetically-biased formalism. 

\subsection{Magnetically Biased Belinfante's spin angular momentum}\label{Appendix:magnetically_biased_spin_dipole}
The electric vector potential can be from 
\begin{equation}
\begin{split}
\bm{C}^{\perp} & = \epsilon_0 \int \frac{d^3 \bm{r}'}{4 \pi |\bm{r} -\bm{r}'|} \frac{\partial \bm{B}(\bm{r}',t)}{\partial t} \\
& = \frac{\epsilon_0}{4 \pi} \left( \frac{\mu_0}{4 \pi}\right) \int \frac{d^3 \bm{r}'}{ |\bm{r} -\bm{r}'|} \left[- \frac{\dot{\bm{m}} - 3 \hat{\bm{r}}' \left(\hat{\bm{r}}' \cdot \dot{\bm{m}} \right)}{r'^3} + \frac{8 \pi}{3} \dot{\bm{m}} \, \delta\left( \bm{r}'\right) \right] \\ 
& = \epsilon_0  \left( \frac{\mu_0}{4 \pi}\right) \frac{\dot{\bm{m}}}{r}
\end{split} 
\end{equation}
The magnetically-biased spin density in the near-field is 
\begin{equation}
    \begin{split}
    \bm{s}_{\rm{m}} & = \bm{B} \times \bm{C}^\perp - \left( \bm{B} \cdot \nabla \right) \left(\bm{r} \times \bm{C}^\perp \right) \\
    & = \bm{B} \times \bm{C}_\perp - \bm{B} \times \bm{C}_\perp - B_i \epsilon_{jkl} r_k \partial_i C^{\perp}_l \\ 
    & = - \frac{\epsilon_0}{4 \pi} \left(\frac{\mu_0}{4 \pi} \right) \left(\hat{\bm{r}} \cdot \bm{B} \right) \frac{\left(\hat{\bm{r}} \times \dot{\bm{m}} \right)}{r} \\
    & = -\frac{\epsilon_0}{4 \pi} \left(\hat{\bm{r}} \cdot \bm{B} \right) \bm{E} \\
    & = 2 \epsilon_0 \left(\frac{\mu_0}{4 \pi}\right)^2 \, \frac{(\hat{\bm{r}} \times \dot{\bm{m}}) (\hat{\bm{r}} \cdot \bm{m})}{r^4}  + \epsilon_0 \left(\frac{\mu_0}{4 \pi} \right)^2\frac{8 \pi}{3} \, \frac{(\hat{\bm{r}} \times \dot{\bm{m}}) (\hat{\bm{r}} \cdot \bm{m})}{r} \, \delta\left( \bm{r} \right)
    \end{split}
    \label{eq:spin_density_precessing_dipole}
\end{equation}
where 
\begin{equation}
  \bm{B} \times \bm{C}^\perp = - \epsilon_0 \left(\frac{\mu_0}{4 \pi}\right)^2 \frac{\bm{m} \times \dot{\bm{m}}}{r^4} +3 \epsilon_0 \left(\frac{\mu_0}{4 \pi}\right)^2 \, \frac{(\hat{\bm{r}} \times \dot{\bm{m}}) (\hat{\bm{r}} \cdot \bm{m})}{r^4}  + \epsilon_0 \left(\frac{\mu_0}{4 \pi} \right)^2\frac{8 \pi}{3} \, \frac{(\hat{\bm{r}} \times \dot{\bm{m}}) (\hat{\bm{r}} \cdot \bm{m})}{r} \, \delta\left( \bm{r} \right)   
\end{equation}
In exponential notation it can be expressed as: 
\begin{equation}
\bm{s}_{\rm{m}} =\epsilon_0 \left(\frac{\mu_0}{4 \pi} \right)^2 \left[ 2 \,  \frac{\, m_0 \, \omega \, \mathfrak{\xi}(x,y,z,t)}{r^5} + \frac{8 \pi}{3} \,\frac{\, m_0 \, \omega \, \mathfrak{\xi}(x,y,z,t)}{r^2} \delta(\bm{r})\right] \left[\Re \left(z \left( \hat{\bm{x}} -i \hat{\bm{y}} \right) e^{i \omega t} \right)  + \Im\left( \left(y -i x \right) e^{i \omega t} \right) \hat{\bm{z}} \right]
\end{equation}
where we used the relation 
\begin{equation}
    \begin{split}
    \left(\bm{n} \times \dot{\bm{m}}\right) \left(\bm{n} \cdot \bm{m} \right) & = m_0 b_0 \,  \xi(x,y,z,t) \left[ \frac{z}{r} \Re \left(z \left( \hat{\bm{x}} -i \hat{\bm{y}} \right) e^{i \omega t} \right)  + \Im\left( \left(y -i x \right) e^{i \omega t} \right) \hat{\bm{z}} \right]
    \end{split}
\end{equation}
and the function $ \xi(x,y,z,t)$ is defined as 
\begin{equation}
\xi(x,y,z,t)  =  \frac{x}{r} m_0 \cos(\omega t)+ \frac{y}{r} m_0 \sin(\omega t) +  b_0 \frac{z}{r} 
\end{equation}

\section{Electrically-biased angular momentum density}
The electric field and the magnetic vector potential of a magnetic dipole in the near-field regime are given by \cite{Zangwill2013}:
\begin{align}
\bm{E} &= \frac{\mu_0}{4\pi} \frac{\hat{\bm{r}} \times \dot{\bm{m}}}{r^2}, \\
\bm{A}^\perp &= \frac{\mu_0}{4\pi} \frac{\hat{\bm{r}} \times \bm{m}}{r^2}.
\end{align}

Since there is no charge density $\rho = 0$ in this system \cite{Stewart2005May}, no orbital angular momentum is expected. Surprisingly, the term associated with electrically-biased orbital angular momentum,
\begin{equation}
\epsilon_0 \bm{E}_i (\bm{r} \times \nabla) A^\perp_i = \epsilon_0 \left(\frac{\mu_0}{4\pi}\right)^2 \frac{(\hat{\bm{r}} \cdot \bm{m})(\hat{\bm{r}} \times \dot{\bm{m}})}{r^4},
\end{equation}
exhibits spin-like rather than orbital character. This reveals that the electrically-biased formalism mixes spin and orbital contributions.

%The complete electrically-biased spin density evaluates to:
%\begin{equation}
%\bm{s}_{\mathrm{e}} = \epsilon_0 \left(\frac{\mu_0}{4\pi}\right)^2 \frac{(\hat{\bm{r}} \cdot \bm{m})(\hat{\bm{r}} \times \dot{\bm{m}})}{r^4},
%\end{equation}
the complete electrically-biased  total angular momentum:
\begin{equation}
\bm{j}_{\rm{bel}} = 2\epsilon_0 \left(\frac{\mu_0}{4\pi}\right)^2 \frac{(\hat{\bm{r}} \cdot \bm{m})(\hat{\bm{r}} \times \dot{\bm{m}})}{r^4}.
\end{equation}
The key distinction lies in the mixing of canonical forms: while the magnetically-biased formalism maintains clean separation between spin and orbital components, the electrically-biased version inherently mixes them.

\section{Frequency decomposition}\label{appendix:frequency_decomposition}
The instantaneous spin can be in general decomposed into its frequency time harmonic components 
\begin{equation}
    \bm{S}\left( t\right) = \bm{S}_0 + \bm{S}_1 \cos(\omega t) + \bm{S}_1 \sin(\omega t) + \bm{S}_2 \cos(2 \omega t) + \bm{S}_2 \sin(2   \omega t) + \cdot \cdot \cdot
\end{equation}
 The constant term $\mathbf{S}_{0}$ is simply the time-average of the spin. If $\mathbf{S}_{0}\neq\mathbf 0$, the field carries a nonzero net (time-averaged) spin angular momentum. For example, a circularly polarized rotating dipole will produce a DC spin along the rotation axis (analogous to a photon’s helicity). In contrast, a linearly oscillating dipole has $\mathbf{S}_{\rm 0}= \mathbf 0$ (its time-average spin vanishes) even though $\mathbf{S}(t)$ may swing back and forth instantaneously. The presence or absence of $\mathbf{S}_{0}$ thus determines whether there is a net angular momentum flux.

The terms with $\omega$ come from spin oscillating at the driving frequency, and the $2\omega$ terms (and higher harmonics) arise from nonlinear combinations of the fields. Similar decomposition can be made for the spin density. 

The spin density in the the case of precessing magnetic dipole consists of three frequencies $\left( 0, \omega, 2 \omega \right)$. This can be shown by as following: 

Write the unit observation vector as $\hat{\mathbf r}=(x,y,z)$ (fixed in space for a chosen observation point). Then
\[
\hat{\mathbf r}\cdot\mathbf m
= m_0\bigl(x\cos(\omega t)+y\sin(\omega t)\bigr)+b_0 z,
\]
\[
\hat{\mathbf r}\times\dot{\mathbf m}
= m_0\omega\bigl(-z\cos(\omega t),\,-z\sin(\omega t),\,x\cos(\omega t)+y\sin(\omega t)\bigr),
\]
where the cross product is evaluated componentwise. Multiplying these factors componentwise yields
\begin{equation}
    \begin{split}
        s_x & =  2 \epsilon_0 \left( \frac{\mu_0}{4 \pi}\right)^2 \left[ -m_0^2\omega\, z\cos(\omega t)\bigl(x\cos(\omega t)+y\sin(\omega t)\bigr)
- m_0\omega b_0\, z^2\cos(\omega t) \right] \left(\frac{1}{r^5} + \frac{4 \pi}{3} \delta\left( \bm{r} \right) \right)\\
s_y & = 2 \epsilon_0 \left( \frac{\mu_0}{4 \pi}\right)^2 \left[-m_0^2\omega\, z\sin(\omega t)\bigl(x\cos(\omega t)+y\sin(\omega t)\bigr)
- m_0\omega b_0\, z^2\sin(\omega t) \right] \left(\frac{1}{r^5} + \frac{4 \pi}{3} \delta\left( \bm{r} \right) \right)\\
s_z & = 2 \epsilon_0 \left( \frac{\mu_0}{4 \pi}\right)^2 \left[ m_0^2\omega\bigl(x\cos(\omega t)+y\sin(\omega t)\bigr)^2
+ m_0\omega b_0\, z\bigl(x\cos(\omega t)+y\sin(\omega t)\bigr)\right] \left(\frac{1}{r^5} + \frac{4 \pi}{3} \delta\left( \bm{r} \right) \right)
    \end{split}
\end{equation}

Expanding the trigonometric products and using standard trigonometric identities give 
\begin{equation}
\begin{split}
s_x &= -\epsilon_0 \left( \frac{\mu_0}{4 \pi}\right)^2 
m_0 \omega z 
\biggl[ m_0 x + 2 b_0 z \cos(\omega t) 
       + x m_0 \cos(2 \omega t) + y m_0 \sin(2 \omega t) \biggr]
\left( \frac{1}{r^5} + \frac{4 \pi}{3} \delta\!\bigl( \bm{r} \bigr) \right) \\[6pt]
s_y &= -\epsilon_0 \left( \frac{\mu_0}{4 \pi}\right)^2 
m_0 \omega z 
\biggl[ m_0 y + 2 b_0 z \sin(\omega t) 
       - y m_0 \cos(2 \omega t) - x m_0 \sin(2 \omega t) \biggr]
\left( \frac{1}{r^5} + \frac{4 \pi}{3} \delta\!\bigl( \bm{r} \bigr) \right) \\[6pt]
s_z &= -\epsilon_0 \left( \frac{\mu_0}{4 \pi}\right)^2 
m_0 \omega 
\biggl[ 
 m_0\!\left( x^2 + y^2 \right) 
 + 2 b_0 z \bigl(x \cos(\omega t) + y \sin(\omega t) \bigr) \\[2pt]
&\quad\quad\;\; 
 + m_0 \bigl( x^2 - y^2 \bigr) \cos(2 \omega t) 
 + 2 m_0 x y \sin(2 \omega t) 
\biggr]
\left( \frac{1}{r^5} + \frac{4 \pi}{3} \delta\!\bigl( \bm{r} \bigr) \right)
\end{split}
\end{equation}

One sees that terms proportional to $m_0^2$ contain both a time-independent (DC) contribution and a $2\omega$ contribution, whereas terms proportional to $m_0 b_0$ oscillate at the fundamental frequency $\omega$. Therefore the full local spin density generically contains components at $0$, $\omega$ and $2\omega$. At a fixed spatial point the spin is time dependent and typically traces an oscillatory (often elliptical) path whose orientation and sense depend on the coordinates $(x,y,z)$. If one parameter dominates (for example, if $b_0\gg m_0$ or vice versa), the motion may appear close to a single-frequency rotation at $\omega$, but generically it is a superposition of multiple frequency components.

\section{Spin density distributions for precessing magnetic dipole}\label{appendix:spin_density}

\begin{figure*}
\includegraphics[width = 0.99 \textwidth]{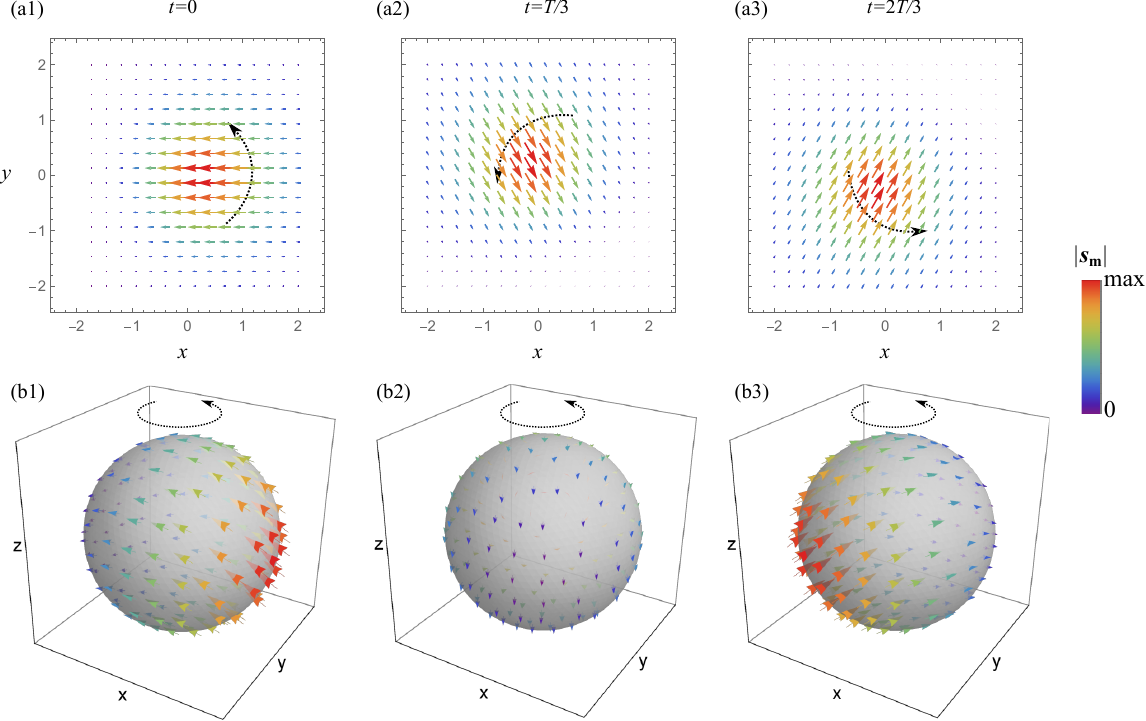}
\caption{
%(a1-a3) Trajectories of local near-field spin density vector tips on a sphere of unit radius around the precessing point dipole (black curves) with $b_0 = m_0$ (b2). 
(a1-a3) $x$ and $y$ components of spin density of the electromagnetic fields from a precessing magnetic dipole in the horizontal cross-section at $z=\text{const}$ at three time instances inside the precession period: $t=0$, $T/3$, and $2 T/3$. 
(b1-b3) Local electromagnetic spin density distributions for the same time instances at the unit sphere. 
%The black dotted arrows in (d-e) act as a guide for the direction of solid-like rotation of the spin density distribution synchronous with dipole precession. 
%
%
%
%Upper 
%Plot shows $x$ and $y$ components of spin density in the two-dimensional cross-section of the distribution at $z=2$ plane by arrows, with $z$ component encoded by color. 
%Lower plots show ... (to be inserted). 
%Parameters: $b_0 = m_0$. 
%
%Proposed experimental setup for a meta-atom
}
\label{figS:spin_precession}
\end{figure*}

Figure~\ref{figS:spin_precession}(a,b) illustrates the spin density components of the precessing magnetic dipole ($b_0 = m_0$) in the two-dimensional cross-section at the $z = \text{const}$ plane [panels (a1-a3)], as well as on the unit sphere [panels (b1-b3)]. 
As expected in teh near field, the local structure of spin density follows dipole precessional motion as a solid structure.

\section{Total EM spin from a precessing magnetic dipole}\label{appendix:total_spin}
As we proved before  
\[
\mathbf s_m(\mathbf r,t)
= \mathbf B(\mathbf r,t)\times \mathbf C_\perp(\mathbf r,t)
- (\mathbf B\!\cdot\!\nabla)\bigl(\mathbf r\times \mathbf C_\perp(\mathbf r,t)\bigr).
\]
Define the total instantaneous quantity
\[
\mathbf S_m(t)
= \int_{\mathbb R^3} \mathbf s_m(\mathbf r,t)\,{\rm d}^3r.
\]
Using the standard identity (integration by parts in index form),
\begin{equation}
    \int (\bm{B} \cdot \bm{\nabla}) \bm{G} \,{\rm d}^3r
= \int \bm{\nabla} \cdot\!(\bm{B} \bm{G})\,{d}^3r
- \int (\bm{\nabla}  \cdot \bm{B}) \bm{G}  d^3r,
\end{equation}
where $\bm{G} =\bm{r} \times \bm{C}^\perp$, we obtain
\begin{equation}
\int (\mathbf B\!\cdot\!\nabla)\bigl(\mathbf r\times \mathbf C_\perp\bigr)\,{\rm d}^3r
= \int \nabla\!\cdot\!\bigl[\mathbf B(\mathbf r\times \mathbf C_\perp)\bigr]\,{\rm d}^3r
- \int (\nabla\!\cdot\!\mathbf B)\,\bigl(\mathbf r\times \mathbf C_\perp\bigr)\,{\rm d}^3r,
\end{equation}
Maxwell’s equation $\nabla\!\cdot\!\mathbf B=0$ eliminates the last volume term. The first term reduces to a surface integral,
\begin{equation}
    \int (\mathbf B\!\cdot\!\nabla)\bigl(\mathbf r\times \mathbf C_\perp\bigr)\,{\rm d}^3r
= \int_{\partial V} \bigl[\mathbf B(\mathbf r\times \mathbf C_\perp)\bigr]\!\cdot\!{\rm d}\mathbf \Gamma,
\end{equation}
evaluated at spatial infinity. If the physical fields decay sufficiently rapidly so that 
$\mathbf B(\mathbf r\times \mathbf C_\perp)$ falls faster than $1/r^{2}$, this surface integral vanishes. Consequently,
\begin{equation}
\mathbf S_m(t)
= \int_{\mathbb R^3} \mathbf B(\mathbf r,t)\times \mathbf C_\perp(\mathbf r,t)\,{\rm d}^3r,
\label{eq:spin_dipole_general}
\end{equation}
showing that the second term in $\mathbf s_m$ contributes only a pure divergence whose integral is zero. The explicit $\mathbf r$ in the original density does not affect the total quantity, confirming that $\mathbf S_m(t)$ is independent of the choice of origin.

By direct integration of Eq.(\ref{eq:spin_density_precessing_dipole}) or using Eq.(\ref{eq:spin_dipole_general}), one find that the spin of the precessing dipole in the near field zone is, 
\begin{equation}
\bm{S}_\text{m}(t) = \frac{4}{3} \epsilon_0\left(\frac{\mu_0}{4\pi}\right)^2 \frac{\bm{m} \times \dot{\bm{m}}}{r_0} \\
\end{equation}
where $r_0$ is the size of the dipole. Hence, EM spin in the near field is characteristic quantity of the fields as it is a function only of the magnetic momemt which describe the near-field. 

%\section{Canonical spin angular momentum of a precessing magnetic dipole}

\end{document}